\documentclass[pdflatex,sn-mathphys-num]{sn-jnl}

\usepackage[utf8]{inputenc}
\usepackage{siunitx}
\usepackage{multirow}
\usepackage{amsmath}
\usepackage{braket}
\usepackage{colortbl}
\usepackage{hyperref}

\DeclareSIUnit\angstrom{\text {Å}}

\begin{document}

\title{Seeing the forbidden: overcoming optical selection rules through nanophotonic integration}

\author*[1,2]{\fnm{Alex H.} \sur{Rubin}}\email{ahrubin@ucdavis.edu}

\author[3]{\fnm{Vytautas} \sur{Žalandauskas}}

\author[1]{\fnm{Pranta} \sur{Saha}}

\author[1,2]{\fnm{Rubek} \sur{Poudel}}

\author[3]{\fnm{Aurora} \sur{Teien}}

\author[1,2]{\fnm{Liam} \sur{Hofmann}}

\author[1,2]{\fnm{Nathan R.} \sur{Gonzalez}}

\author[4]{\fnm{Scott} \sur{Dhuey}}

\author[3]{\fnm{Marianne} \sur{Etzelm\"uller Bathen}}

\author*[1]{\fnm{Marina} \sur{Radulaski}}\email{mradulaski@ucdavis.edu}

\affil[1]{\orgdiv{Department of Electrical and Computer Engineering}, \orgname{University of California, Davis}, \orgaddress{\city{Davis}, \state{CA}, \postcode{95616}, \country{United States of America}}}

\affil[2]{\orgdiv{Department of Physics and Astronomy}, \orgname{University of California, Davis}, \orgaddress{\city{Davis}, \state{CA}, \postcode{95616}, \country{United States of America}}}

\affil[3]{\orgdiv{Department of Physics/Centre for Materials Science and Nanotechnology}, \orgname{University of Oslo}, \orgaddress{\postcode{0316}, \city{Oslo}, \country{Norway}}}

\affil[4]{\orgdiv{The Molecular Foundry}, \orgname{Lawrence Berkeley National Laboratory}, 
\orgaddress{\city{Berkeley}, \state{CA}, \postcode{94720}, \country{United States of America}}}

\date{\today}

\abstract{
Optically addressable spin defects in silicon carbide, including the neutral divacancy (VV$^{0}$) and the negative nitrogen–vacancy (NV$^{-}$), are among leading building blocks of solid-state quantum technologies.
Integrating these defects into photonic structures such as nanopillars improves photon collection efficiency, but the consequences extend further.
We show that the sub-wavelength geometry of nanopillars drastically modifies the local electromagnetic environment, providing optical access to defect transitions that are otherwise suppressed by selection rules in bulk material.
Using low-temperature photoluminescence spectroscopy, we observe that emission from the PL3 divacancy, which is nearly absent in planar devices, becomes pronounced in nanopillars owing to a polarization transformation of the excitation field within the pillar.
We further leverage the orientation-dependent collection of nanopillars to resolve the origin of previously ambiguous spectral lines. In particular, the NV4$'$ feature displays the signal enhancement expected for axially oriented NV$^{-}$ centres, consistent with assignment to a higher excited state of the $kh$ defect configuration.
Our results establish nanophotonic integration as a symmetry-sensitive probe that can both activate nominally dark transitions and identify the dipole character of poorly understood defect states.
}

\maketitle

\section{Introduction}
The rapidly advancing field of quantum optical technologies promises to revolutionize secure communications, metrology, and distributed information processing.
Realizing this promise requires scalable hardware platforms and robust single-photon sources and spin-photon entangling interfaces, which has driven intense interest in wide-bandgap semiconductors such as 4H-silicon carbide (4H-SiC).
This material hosts a variety of deep-level point defects, including the negatively charged nitrogen-vacancy ($\mathrm{V_{Si}N_{C}}^{-}$, NV$^{-}$) \cite{PhysRevB.92.064104, PhysRevB.94.060102, Mu2020, Wang2020, zhigulin2026couplingnitrogenvacancycenters} and the neutral divacancy ($\mathrm{V_{Si}V_{C}}^{0}$, VV$^{0}$) \cite{koehl_room_2011, christle_isolated_2015, anderson_five-second_2022} centers, which combine long-lived spin coherence with bright optical emission.
To meet the demands for high-rate single-photon sources, one well-established photonic integration technique is to fabricate nanopillars, which direct the emission of embedded color centers out of plane, yielding optical signal enhancement of an order of magnitude or more \cite{radulaski2017scalable, Norman2025}.

The optical properties of defects of this type are governed by their local point-group symmetry, which imposes strict selection rules defining the allowed optical transitions.
For 4H-SiC wafers, which are conventionally grown along the $c$-axis, a standard confocal microscopy experiment on a planar substrate sends the excitation light along the $c$-axis with polarization confined to the basal plane.
Transitions whose dipole moment lies along $c$ are therefore effectively forbidden under this excitation.
Nanophotonic integration perturbs this picture in two underexplored ways.
First, we have previously shown that the collection efficiency of a nanopillar is sensitive to the orientation of the defect's optical dipole moment, allowing the pillar to fingerprint distinct crystallographic configurations of a given emitter \cite{Norman2025}.
Second, the sub-wavelength geometry restructures the local optical field, generating substantial $c$-axis polarization components within the pillar that are nearly absent in bulk and unlocking transitions otherwise suppressed by selection rules.

\begin{figure*}
\includegraphics[width=1.00\textwidth]{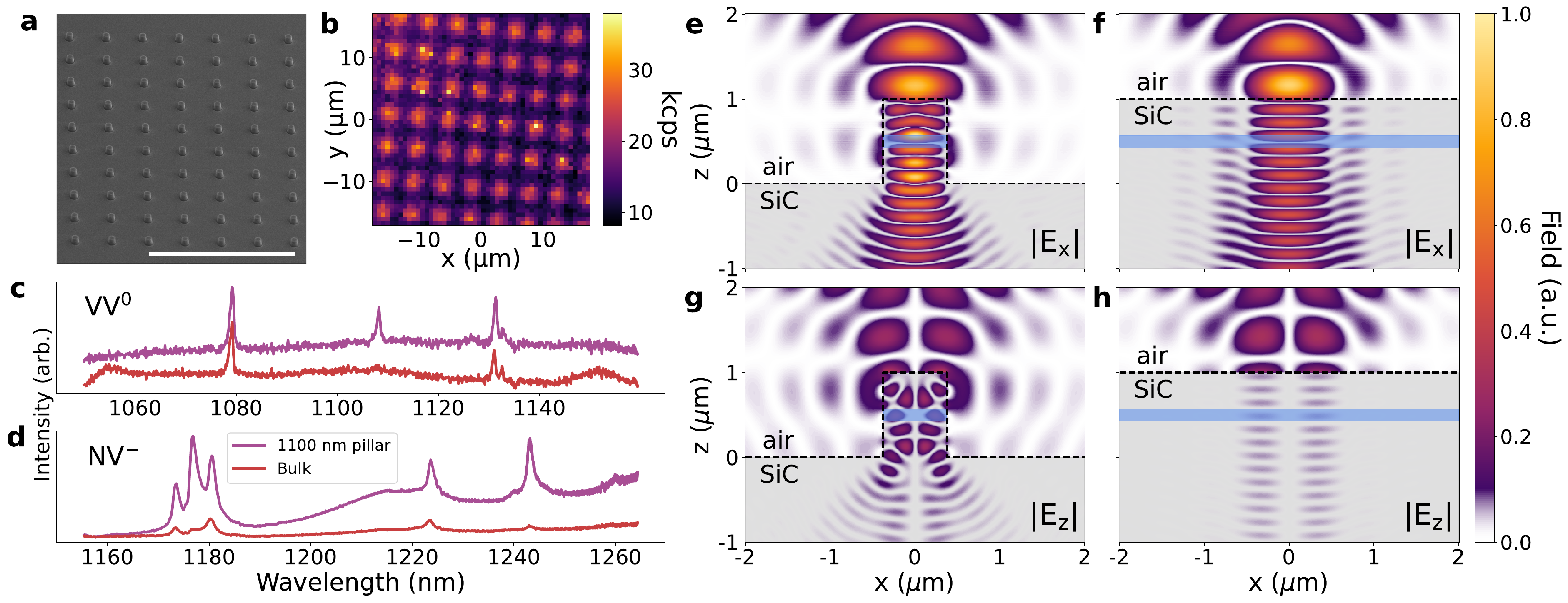}
  \caption{
  {\boldmath\textbf{The nanopillar enables prohibited and enhances allowed optical transitions of VV$^{0}$ and NV$^{-}$ defects in 4H-SiC, compared to the bulk, by altering the polarization of the excitation laser.}}
  \textbf{a} An SEM image of an array of 850~nm diameter and \SI{1}{\micro\meter} height 4H-SiC nanopillars. The scale bar represents \SI{20}{\micro\meter}.
  \textbf{b} 2D photoluminescence map of an array of 600~nm diameter nanopillars taken via scanning confocal microscopy using 785~nm excitation and SNSPD detection.
  \textbf{c}, \textbf{d} Examples of photoluminescence spectra of VV$^{0}$ and NV$^{-}$ defects taken from 1100~nm diameter nanopillars and from unetched bulk areas of the sample.
  Each spectrum is an average of 10 individual spectra taken from nominally identical areas.
  \textbf{e}--\textbf{h} The FDTD simulations of the electrical field components perpendicular ($\mathrm{E}_x$) and parallel ($\mathrm{E}_z$) to the crystal axis ($c$) of the 785~nm excitation light inside a 750~nm diameter nanopillar (\textbf{e} and \textbf{g}) and bulk silicon carbide (\textbf{f} and \textbf{h}).
  The blue-shaded section indicates the region of nitrogen ion implantation as determined by the Stopping and Range of Ions in Matter (SRIM) simulations spanning the projected range of NV$^{-}$ centers to $500 \pm 70$~nm depth.
  }
  \label{fig:FDTD}
\end{figure*}

Here we demonstrate two consequences of this interplay between nanophotonic geometry and defect symmetry, supported by the first-principles calculations of the transition dipole moments of the NV$^{-}$ and VV$^{0}$ centers in 4H-SiC and a group-theoretical symmetry analysis of their optical transitions.
First, we show that the PL3 divacancy center, which is nearly invisible when using confocal microscopy on bulk samples due to selection rules requiring $c$-axis polarization, becomes prominently visible when integrated into a nanopillar.
Second, we exploit the orientation sensitivity of nanopillar collection efficiency to explore the NV4$'$ spectral line, which appears in nearly all 4H-SiC NV$^{-}$ spectra but whose origin has been unclear \cite{ivanov2023near}.
Our theoretical analysis shows that this line originates from a higher excited state of the $kh$ configuration of the NV$^{-}$ center which arises from the lower symmetry of the basal configurations, and that its emission is dipole-forbidden for excitation light polarized along the $c$-axis, similar to the optical transitions of axial NV$^{-}$ centers.
We show that the NV4$'$ line exhibits a signal enhancement in nanopillars typical of axial NV$^{-}$ centers, which experimentally corroborates our symmetry-based argument.
Temperature-dependent photoluminescence measurements provide additional confirmation: NV4$'$, as well as the PL3$'$ divacancy line (which arises for similar reasons), exhibit negative thermal quenching at low temperatures, a characteristic signature of emission from a thermally populated higher-lying excited state \cite{Shibata_1998, PhysRevB.82.115207}.
Together, these results establish that nanophotonic integration can serve both as a method of engineering access to color center transitions for practical applications and as a tool for clarifying the details of their internal electronic structure.

\section{Results}

\subsection{Electronic structure and symmetry analysis \label{sec:structure_and_symm}}

\begin{figure*}
\includegraphics[width=1.00\textwidth]{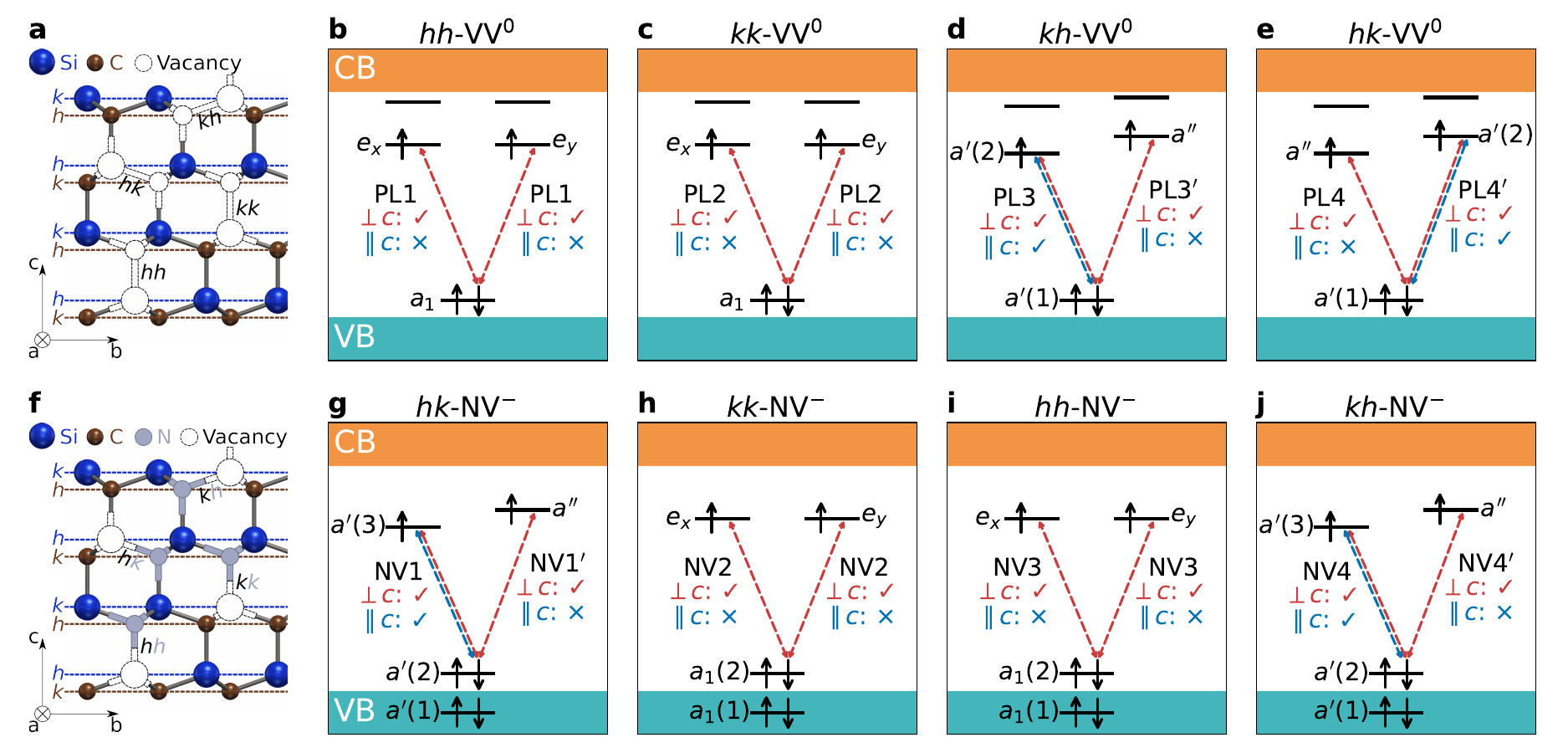}
    \caption{
    {\boldmath\textbf{Crystal and electronic structure of VV$^0$ and NV$^-$ defects in 4H-SiC}}.
    \textbf{a}~Ball-and-stick representation of divacancy defects. Blue, brown, and white spheres denote silicon, carbon, and vacancy sites, respectively. The labels $h$ and $k$ indicate the hexagonal and quasi-cubic lattice sites in 4H-SiC.
    \textbf{b}--\textbf{e}~Kohn--Sham single-particle defect level diagrams for the four nonequivalent VV$^0$ configurations ($hh$, $kk$, $kh$, $hk$) and optical transitions associated with the PL1, PL2, PL3/PL3$'$, and PL4/PL4$'$ lines, respectively.
    \textbf{f}~Ball-and-stick representation of nitrogen-vacancy defects. Gray spheres additionally denote nitrogen atoms.
    \textbf{g}--\textbf{j}~Kohn--Sham single-particle defect level diagrams for the four nonequivalent NV$^-$ configurations ($hk$, $kk$, $hh$, $kh$) and optical transitions associated with the NV1/NV1$'$, NV2, NV3, and NV4/NV4$'$ lines, respectively.
    In all defect level diagrams, upward and downward black arrows on each level indicate spin-majority and spin-minority occupation, respectively. Shaded regions denote the valence band (VB, teal) and conduction band (CB, orange).
    Dashed arrows represent optically allowed transitions to excited states, with arrow color indicating the polarization selection rules: red corresponds to $\mathrm{E} \perp c$,  and blue corresponds to $\mathrm{E} \parallel c$.
    }
\label{fig:KS_levels}
\end{figure*}

In the 4H-SiC polytype, Si--C bilayers stack along the crystallographic $c$-axis (the $[0001]$ direction), with each atomic site occupying either a hexagonal (\textit{h}) or a quasi-cubic (\textit{k}) lattice configuration. As a result, four nonequivalent configurations of both the divacancy (VV) and the nitrogen-vacancy (NV) complex arise, denoted as $hh$, $kk$, $hk$, and $kh$ according to the local symmetry (see Fig.~\ref{fig:KS_levels}\textbf{a},\textbf{f}). In this work, we investigate the neutral charge state of the divacancy (VV$^0$) and the negative charge state of the nitrogen-vacancy center (NV$^-$). In the ground state, the axial configurations ($hh$ and $kk$), which are confined to the $c$-plane of the 4H-SiC crystal, exhibit $C_{3v}$ symmetry, whereas the basal configurations ($kh$ and $hk$) lie in the basal plane and possess the reduced $C_{1h}$ symmetry. These symmetries determine the orientation of each configuration's transition dipole moment (TDM) relative to the crystallographic axes, which in turn governs the optical selection rules.

Fig.~\ref{fig:KS_levels} shows the ground state Kohn--Sham single-particle level diagrams for the four configurations of both the VV$^{0}$ and NV$^{-}$ defects. Dashed arrows denote the symmetry-allowed optical transitions to the first and second excited states. The arrow color encodes the polarization selection rules: red corresponds to transitions allowed under $\mathrm{E} \perp c$ (i.e., $\mathrm{E}_x$ and $\mathrm{E}_y$), while blue indicates transitions allowed for $\mathrm{E} \parallel c$ (i.e., $\mathrm{E}_z$). Here, $\mathrm{E}$ denotes the electric field polarization of the ZPLs, consistent with the Cartesian components $\mathrm{E}_{x}$ and $\mathrm{E}_{z}$ in Fig.~\ref{fig:FDTD}\textbf{e--h}.

Axial defect configurations introduce an $a_1$ state and a higher lying set of doubly degenerate $e$ states ($e_x$ and $e_y$) derived from carbon dangling bonds within the band gap of 4H-SiC (see Figs.~\ref{fig:KS_levels}\textbf{b},\textbf{c} and \textbf{h},\textbf{i}). These configurations give rise to a $^3\!A_2$ ground state and a doubly degenerate $^3\!E$ excited state. The two resulting optical transitions are degenerate and have a transition dipole moment that lies entirely in the basal plane, rendering them active only for light polarized perpendicularly to the $c$-axis ($\mathrm{E} \perp c$), as indicated by the red dashed arrows. 

In contrast, basal defect configurations possess reduced symmetry which requires a more nuanced analysis. Here, the reduced symmetry lifts the orbital degeneracy and splits the states into irreducible representations $a'$ or $a''$ (see Figs.~\ref{fig:KS_levels}\textbf{d},\textbf{e} and \textbf{g},\textbf{j}).
All basal defect configurations possess a triplet ground state of $^3\!A''$ symmetry. For the $kh$-VV$^{0}$, $hk$-NV$^-$ and $kh$-NV$^-$ configurations, the first excited state also has ${^3\!A''}$ symmetry, whereas for the $hk$-VV$^{0}$ configuration the symmetry group is ${^3\!A'}$. Conversely, the second excited state has ${^3\!A'}$ symmetry for the $kh$-VV$^{0}$, $hk$-NV$^-$, and $kh$-NV$^-$ configurations, while the $hk$-VV$^{0}$ configuration has ${^3\!A''}$ symmetry. Optical transitions between $^3\!A''$ and $^3\!A''$ states impose no parity change, while those between $^3\!A'$ and $^3\!A''$ do, requiring the TDM to transform as $A'$ and  $A''$, respectively. While making no limits to the former (indicated by the red and blue dashed arrows), the latter scenario bounds optical polarization to be perpendicular to the mirror plane of the defect (indicated by the red dashed arrows).
These ZPL selection rules are consistent with previously reported theoretical and experimental studies~\cite{Davidsson2020_TDM,Shafizadeh_2024,Stenlund_2025}. A more detailed analysis of the selection rules for the $C_{3v}$ and $C_{1h}$ point groups is provided in Supplementary Notes~2 and 3.

\begin{figure*}
  \includegraphics[width=1.00\textwidth]{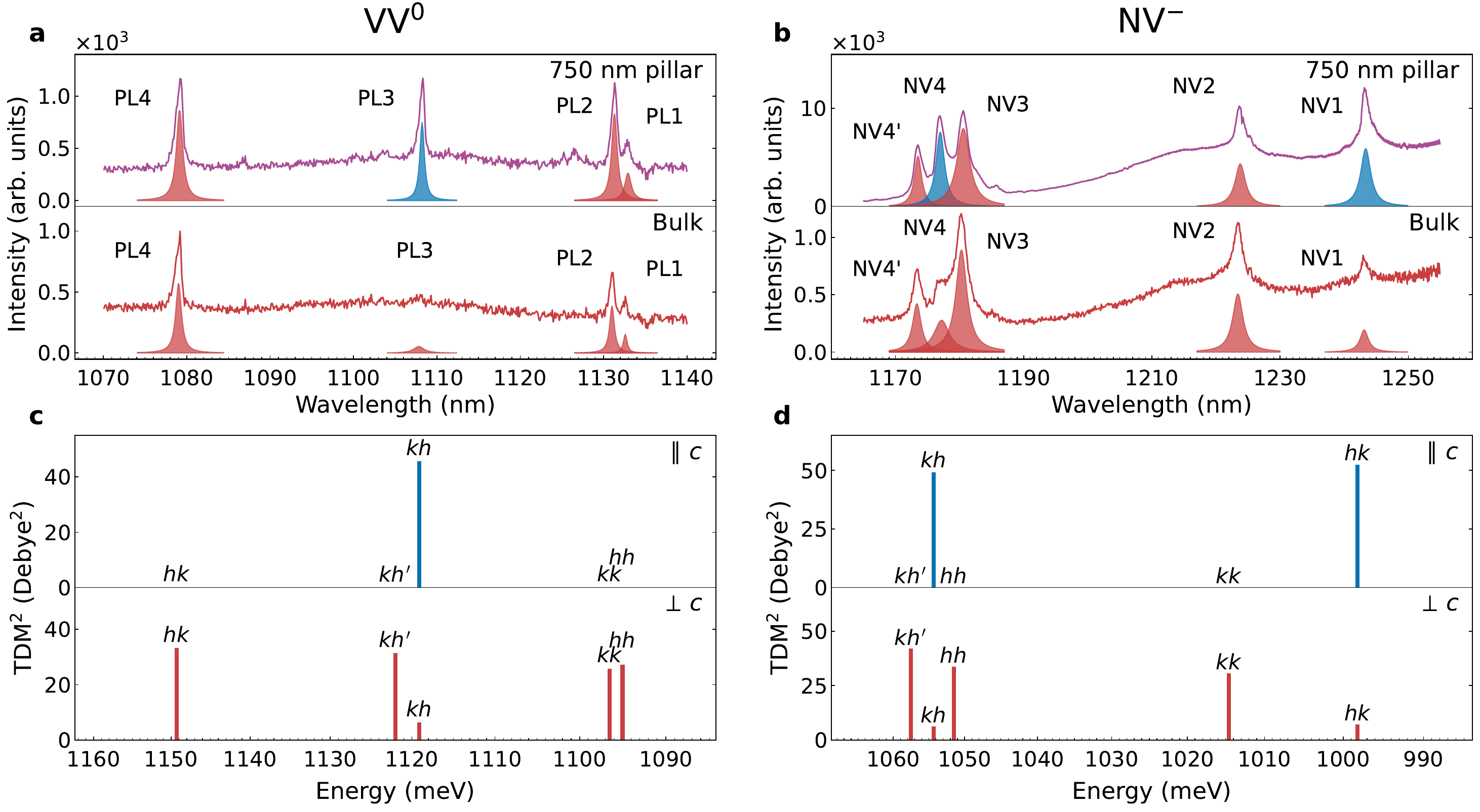}
  \caption{
  {\boldmath\textbf{Photoluminescence spectra and calculated transition dipole moments of VV$^{0}$ and NV$^{-}$ defects in bulk and nanopillar geometries}}.
 Cryogenic photoluminescence spectra of  \textbf{a} divacancy VV$^{0}$ ($\mathrm{V_{Si}V_{C}^{0}}$) and \textbf{b} NV$^{-}$ ($\mathrm{V_{Si}N_{C}^{-}}$) centers collected in unetched bulk material (bottom) and from 750~nm diameter nanopillars (top) at $T=1.6$~K.
  Each spectrum is the average of 10 individual spectra taken from different nanopillars of the same size or different areas of the bulk.
  In both cases the excitation laser at 785~nm was linearly polarized in the plane of the sample, perpendicular to the $c$-axis.
  Defects in bulk are mainly excited by light perpendicularly polarized to the $c$-axis 
  coming directly from the laser; defects in the nanopillars are excited by both the parallel ($\mathrm{E} \parallel c$) and the perpendicular ($\mathrm{E} \perp c$) components of light, due to the pillars transforming laser polarization.
  \textbf{c, d} Calculated squared transition dipole moment (TDM$^2$) values for the ZPL transitions of each type of defect.
  Top and bottom panels show the perpendicular ($\mathrm{E} \perp c$) and parallel ($\mathrm{E} \parallel c$) polarization components along the $c$-axis, respectively. Labels denote the defect configuration, where $kh'$ indicates the second excited state of the $kh$ configuration for both VV$^{0}$ and NV$^{-}$.
  }
  \label{fig:spec_and_TDM}
\end{figure*}

\subsection{Polarization response}
\subsubsection{Appearance of PL3, NV1 and NV4}
The nonequivalent configurations of the VV$^{0}$ and NV$^{-}$ centers give rise to the distinct zero-phonon lines seen in their photoluminescence emission.
Historically these are labeled PL1 -- PL4 for the VV$^{0}$ and NV1 -- NV4 for the NV$^{-}$.
Fig.~\ref{fig:spec_and_TDM}\textbf{a},\textbf{b} shows photoluminescence spectra collected at 1.6~K from 750~nm diameter nanopillars and from unetched bulk regions of the same sample.
In the bulk spectra, a large variation in the brightness of the individual peaks can be observed: the PL3, NV1 and NV4 peaks are significantly dimmer than the others.
Strikingly, the PL3 peak is barely visible at all.

To understand this, we turn to the transition dipole moments calculated from first principles, shown in Fig.~\ref{fig:spec_and_TDM}\textbf{c},\textbf{d}.
For the PL3 line ($kh$-VV$^0$), the TDM is oriented predominantly parallel to the $c$-axis, so for a sample grown along $c$ the PL3 transition is effectively dark under standard confocal excitation, where light propagates along $c$ and is therefore polarized perpendicular to it.
The other VV$^{0}$ configurations, by contrast, have TDMs predominantly oriented in the basal plane, making them readily accessible under standard excitation schemes.
A similar situation holds for the NV1 and NV4 lines, explaining their relative dimness in the bulk.

The nanopillar geometry breaks this constraint.
As can be seen in Fig.~\ref{fig:FDTD}\textbf{e},\textbf{g}, when the pillar is illuminated with in-plane polarized light, it restructures the local optical field, generating components along the crystal axis.
Averaged over the nitrogen-implanted region, $|\mathrm{E}_{z}|^2$ inside the pillar is roughly~8.1 times larger than in unetched bulk under identical illumination, while $|\mathrm{E}_{x}|^{2}$ is enhanced by a factor of about~1.7.
The $c$-axis polarization component couples strongly to the PL3 transition dipole, providing optical access which is not available in bulk when using a confocal setup.
The nanopillar functions here as a local polarization converter which unlocks a transition forbidden in bulk by symmetry.

\begin{figure}
\centering
\includegraphics[width=\linewidth]{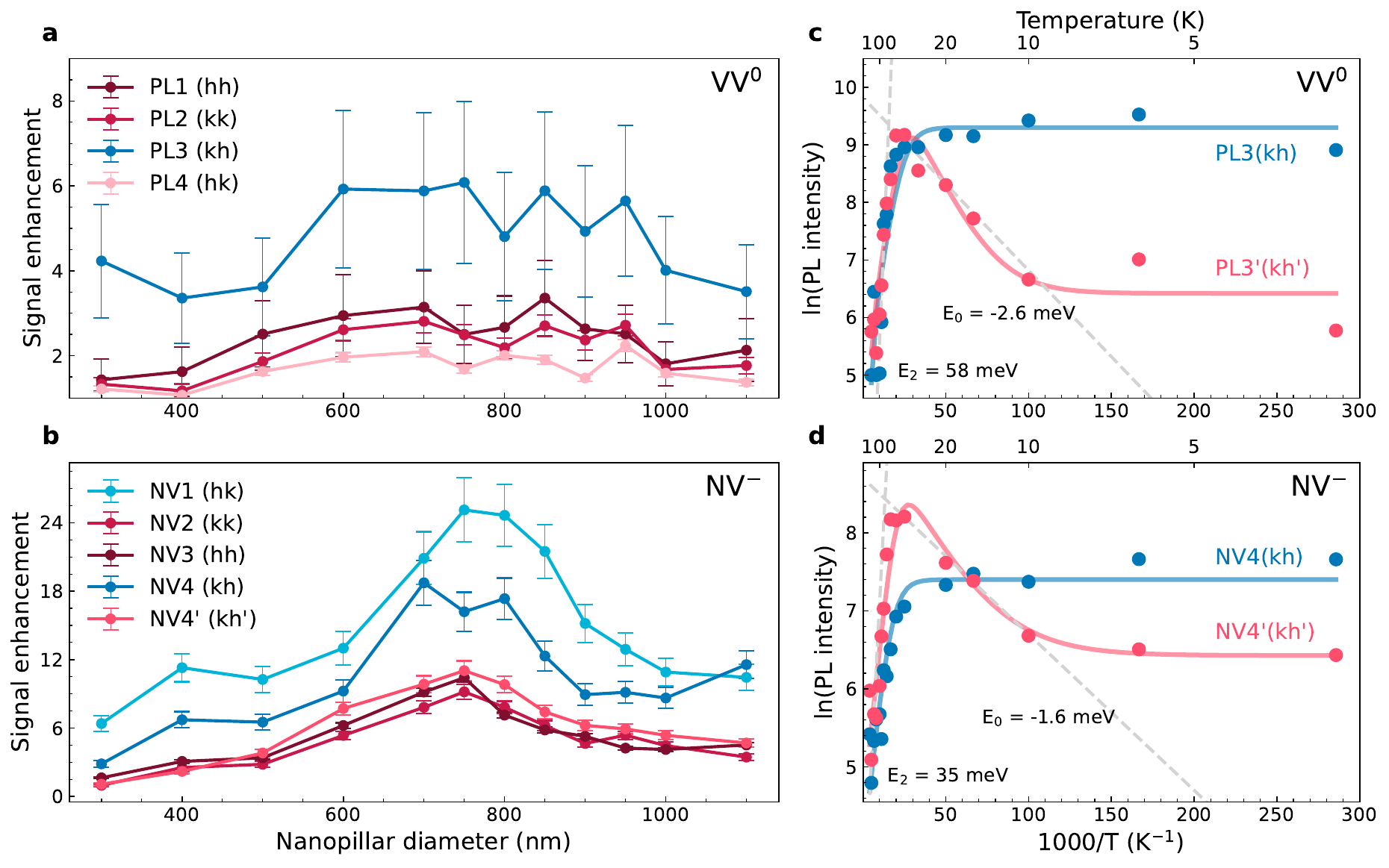}
  \caption{
  {\boldmath\textbf{The VV$^0$ and NV$^-$ signal intensities are altered by pillar integration and temperature variation.}}
  Panels \textbf{a},\textbf{b} were measured at 1.6~K on the nanopillar sample, while \textbf{c},\textbf{d} were measured on a separate unpatterned n-type 4H-SiC sample (see Methods).
  Measured signal in pillars divided by the bulk value as a function of the nanopillar diameter for \textbf{a} each of the four ZPLs of the neutral divacancy (VV$^0$) and \textbf{b} each of the five ZPLs of the negatively charged nitrogen-vacancy (NV$^-$) defects visible from that sample at 1.6~K.
  Arrhenius plots showing the dependence of PL collection temperature of \textbf{c} the PL3 ($kh$) and PL3$'$ ($kh'$) signatures of VV$^0$, and \textbf{d} the NV4 ($kh$) and NV4$'$ ($kh'$) NV$^-$ signatures, in n-type 4H-SiC samples implanted with 300 keV nitrogen ions to a fluence of $1\times10^{13}$~\SI{}{\per\centi\meter\squared}.
  The temperature series measurements were conducted with 852~nm excitation at an incident angle of \SI{27}{\degree} relative to the surface normal.
  The solid curves in \textbf{c},\textbf{d} are asymptotic fits with the activation energy parameters, indicating that PL3$'$ and NV4$'$ experience negative thermal quenching at low temperatures.
  }
  \label{fig:enhancement_tempseries}
\end{figure}

The nanopillar geometry also modifies the outbound collection efficiency of the emitted photons by redirecting emission out of the plane of the sample.
The convolution of these two effects is visible in total signal enhancement factor (defined as the ratio of integrated PL intensity in the pillar to the bulk) as a function of pillar diameter (Fig.~\ref{fig:enhancement_tempseries}\textbf{a},\textbf{b}).
The relative comparison between orientations within each defect family reveals the underlying transition symmetries.
For defects that are efficiently excited in bulk (the axial NV$^{-}$ and VV$^{0}$ centers, as well as PL4), the measured enhancement is driven primarily by the waveguiding of their emission.
However, for PL3, NV1, and NV4, the measured enhancement is a combination of this improved collection and the boosted excitation by the $\mathrm{E}_z$ field component generated within the pillar. 

\subsubsection{Origin of \texorpdfstring{NV4$\,'$ and PL3$\,'$}{NV4' and PL3'} peaks}

Most cryogenic photoluminescence (PL) spectra of NV$^{-}$ centers in 4H-SiC exhibit five ZPLs, including a set of three peaks near 1178~nm.
Initially, the two highest-energy lines (NV4 and NV4$'$) were erroneously assumed to be due to tungsten, and this explanation has lingered in the literature \cite{PhysRevB.94.060102, PhysRevB.98.214113, zhigulin2026couplingnitrogenvacancycenters}.
Other recent reports have observed NV4$'$ without remarking on its relationship to the other ZPLs \cite{PhysRevLett.124.223601, Mu2020}.
However, as discussed in Sec.~\ref{sec:structure_and_symm}, symmetry analysis of the $kh$-NV$^-$ defect predicts the existence of a second excited state $^3\!A'$, giving rise to the NV4$'$ line.
The ${^3\!A'} \rightarrow {^3\!A''}$ transition should be allowed for excitation light polarized $\perp c$ and forbidden for light polarization $\parallel c$, which matches the selection rules for the axial $kk$-NV$^-$ and $hh$-NV$^-$ defects (NV2 and NV3 lines).
Turning to the PL spectra in Fig.~\ref{fig:spec_and_TDM}\textbf{b} we see that the NV4$'$ line is already brightly visible in bulk, consistent with its TDM lying in the basal plane (Fig.~\ref{fig:spec_and_TDM}\textbf{d}).
Indeed, its signal enhancement due to the pillar is quite close to that achieved by the two axial NV$^{-}$ configurations (Fig.~\ref{fig:enhancement_tempseries}\textbf{b}).

For the divacancies, a similar situation holds for the PL3$'$ line, originating from the second excited state $^3\!A'$ of the $kh$-VV$^0$ defect. Like the NV4$'$ transition, the PL3$'$ is predicted by symmetry analysis to obey selection rules similar to the axial defects.
However, unlike the NV4$'$ transition, the $^3\!A'$ second excited state of $kh$-VV$^0$ has a low population at deep cryogenic temperatures, rendering it invisible in our 1.6~K spectra (Fig.~\ref{fig:spec_and_TDM}\textbf{a}), but becoming more pronounced at higher temperatures (Fig.~\ref{fig:enhancement_tempseries}\textbf{c}).
The PL3$'$ peak has often been observed in divacancy PL spectra with a ZPL of 1103~nm, without a detailed discussion of its origin \cite{koehl_room_2011, falk_polytype_2013, son_modified_2022}. 
Both NV4$'$ and PL3$'$ have previously been attributed to a second excited state, however, this assignment lacks experimental validation \cite{Shafizadeh_2024, Shafizadeh_2025}. 
To experimentally confirm the existence of PL3$'$ and NV4$'$, and additionally observe their population dynamics, we therefore collect PL spectra at elevated temperatures and investigate their temperature dependence.

PL spectra were collected from n-type 4H-SiC samples without nanopillars, see Section~\ref{sec:sample_prep}, at different temperatures ranging from cryogenic (3.5~K) to room temperature (300~K).
The resulting Arrhenius plots are shown in Fig.~\ref{fig:enhancement_tempseries}\textbf{c} (PL3, PL3$'$) and ig.~\ref{fig:enhancement_tempseries}\textbf{d} (NV4, NV4$'$), with asymptotic fits and local linear fits whose slopes give activation energies for each temperature-dependent process.
A 852~nm laser excitation was used for the PL measurements with an incident angle of \SI{27}{\degree} relative to the surface normal to excite defects with different polarization dependence. 
The thermal evolution of the PL3 and NV4 lines, associated with the ZPL emission of $kh$-VV$^0$ and $kh$-NV$^{-}$ respectively, exhibit conventional thermal quenching behavior with increasing temperature. In contrast, the PL3$'$ and NV4$'$ lines display negative thermal quenching at low temperatures ($\mathrm{T}<50$~K), characterized by an initial increase in intensity with increasing temperature above 3.5~K.
This behavior is consistent with PL3$'$ and NV4$'$ originating from the higher-lying $^3\!A'$ excited states of $kh$-VV$^0$ and $kh$-NV$^-$, respectively (Fig.~\ref{fig:KS_levels}\textbf{d},\textbf{j}).
Similar negative thermal quenching behavior has been observed for other 4H-SiC defects with higher excited states, such as the silicon vacancy (V$\mathrm{_{Si}}$) \cite{bathen_resolving_2021}. The observation of negative thermal quenching for PL3$'$ and NV4$'$ in Fig.~\ref{fig:enhancement_tempseries}\textbf{c-d} strongly suggests that these peaks stem from the second excited states of $kh$ configurations of the VV$^0$ and NV$^-$ defects, respectively. 

\section{Discussion}
The integration of solid-state defects into nanophotonic structures has historically been motivated by a singular practical objective: maximizing the extraction of indistinguishable single photons into a usable optical mode.
Our findings demonstrate that sub-wavelength photonic devices do considerably more than funnel light.
By restructuring the local electromagnetic environment on a scale comparable to the optical wavelength, they interact with the intrinsic point-group symmetry of embedded defects in ways that can be harnessed as a spectroscopic tool—both for unlocking transitions that are dark in bulk and for disambiguating the dipole character of poorly understood states.

The PL3 spectral line illustrates the first use.
Its transition dipole lies predominantly along the $c$-axis, perpendicular to the polarization of any beam propagating along $c$, rendering it effectively absent from confocal measurements on bulk material.
The pillar circumvents this constraint passively: in-plane polarized light scattering from the dielectric sidewalls develops substantial $\mathrm{E}_z$ components within the pillar volume, supplying precisely the field orientation that the PL3 dipole requires.
A transition that is effectively dark in bulk becomes one of the more prominent features of the nanopillar spectrum, without any modification to the excitation scheme.

The NV4$'$ line illustrates the complementary use: discrimination rather than access.
Although NV4$'$ is visible in bulk,  its microscopic origin has been contested.
The pillar settles the question: NV4$'$ exhibits a diameter-dependent enhancement tracking that of the axial NV$^{-}$ lines rather than the basal ones, indicating a basal-plane transition dipole.
This is exactly the behavior predicted for the $^3\!A'$ higher excited state of the $kh$-NV$^-$ by the symmetry analysis of Section~\ref{sec:structure_and_symm}, and it is difficult to reconcile with an alternate extrinsic impurity origin.
This picture is independently corroborated by the negative thermal quenching of both NV4$'$ and the analogous PL3$'$ line, the signature of emission from a thermally populated higher-lying state.
Together, these provide two independent experimental confirmations that the $kh$ configurations of both defect types possess $^3\!A'$ higher excited states giving rise to secondary spectral lines.

The role of nanophotonic integration here is to supply an experimental handle on predictions that would otherwise be difficult to test directly.
Identifying the microscopic origin of an observed spectral feature has historically been a challenging problem in defect spectroscopy, typically requiring magnetic resonance techniques, single-defect optical measurements, or detailed comparison with first-principles calculations.
The approach demonstrated here adds another lever.
Because the optical enhancement of a ZPL reflects the orientation of its underlying dipole, simple spectroscopic measurements of pillars compared to bulk yield orientation information without requiring single-defect addressing, polarization control, or applied fields.

The nanopillars used in this work were not designed with selection-rule engineering in mind; their geometry was optimized for collection efficiency in the NV$^{-}$ emission band, and the $c$-axis field components responsible for the PL3 visibility are in some sense an incidental byproduct.
Looking ahead, photonic structures purposefully designed to shape the local field could couple to (or selectively avoid) transitions of a given symmetry character far more effectively.
More broadly, the underlying principle is not specific to divacancies, NV$^{-}$ centers, or even to silicon carbide: any defect platform whose optics are governed by symmetry-imposed selection rules is in principle amenable to the same approach.
We anticipate that nanophotonic integration will increasingly take on a dual role in the study of solid-state emitters, serving not only as the standard means of improving photon collection but also as a routine tool for accessing transitions that bulk geometries leave dark and for revealing the structure of the defects themselves.

\section{Methods}

\subsection{Sample preparation}
\label{sec:sample_prep}

The nanopillar sample was fabricated from an HPSI wafer of 4H-SiC purchased from a commercial vendor (Wolfspeed) \cite{Norman2025}.
The substrate was diced into $5\times5$~mm$^2$ pieces, which were then implanted with nitrogen ions (CuttingEdge Ions, LLC) at an energy of 375~keV and a fluence of $10^{14}$~cm$^{-2}$.
The nitrogen implantation has a projected range of 500~nm as estimated by Stopping and Range of Ions in Matter (SRIM) simulations \cite{Ziegler2010}. 
The implanted sample was then annealed at 1050~$^\circ$C in a nitrogen atmosphere for 1~hour.
A 350~nm PMMA layer was applied to the surface, and this was patterned with holes (diameters ranging from 300~nm to 1100~nm) using standard e-beam lithography techniques.
A hard mask consisting of a 5~nm titanium adhesion layer followed by 50~nm of nickel was deposited on top by e-beam evaporation to transfer the circle pattern onto the silicon carbide; the PMMA layer was then lifted off.
The pattern was then etched using inductively coupled SF$_6$ and O$_2$ plasma to produce \SI{1}{\micro\meter} tall pillars.

The sample used for the PL temperature series is from an n-type 4H-SiC wafer, grown \SI{4}{\degree} off-axis with a \SI{10}{\micro\meter} thick epitaxial layer having \SI{1e15}{\per\centi\meter\cubed} free carrier concentration purchased from CREE/Wolfspeed. 
N-implantations were performed in-house at the UiO MiNaLab at an energy of 300~keV and a fluence of \SI{1e13}{\per\centi\meter\squared}, having a projected range of 400~nm according to SRIM. The sample was then exposed to a subsequent annealing step at \SI{1000}{\celsius} for 30~min in an argon atmosphere in a conventional tube furnace. 

\subsection{FDTD simulations}

The simulations were performed using the Ansys Lumerical FDTD solver with a constant refractive index $n_\mathrm{SiC}$ = 2.6.
A 785~nm wavelength Gaussian beam with polarization along the x-axis is injected through a 0.85~numerical aperture using the thin-lens objective method, with the focal point positioned 10~nm below the top surface of the silicon carbide (either the top of the nanopillar or the bulk surface).
To ensure numerical accuracy, we used a uniform mesh with a 10~nm voxel size, and collected the instantaneous field data with a 0.4~fs field-monitor window centered at the excitation pulse maximum. 

\subsection{Cryogenic PL spectroscopy}

The sample was mounted in a 1.6~K optical cryostat (Montana Instruments xp100) fitted with a 100x microscope objective with 0.85 numerical aperture which formed one end of a homebuilt confocal microscope.
A 785~nm laser (ThorLabs MCLS) was used to optically excite the sample.
The collected light was analyzed with a spectrometer consisting of a 750 mm monochromator and a liquid nitrogen-cooled camera (Princeton Instruments PyLoN-IR).

\subsection{First-principles calculations}

The electronic structure calculations for neutral divacancy and negatively charged nitrogen-vacancy defects in 4H-SiC have been performed within the spin-polarized density functional theory (DFT) framework. All DFT calculations were carried out with the Vienna \textit{ab initio} Simulation Package (VASP)~\cite{Kresse_1993,Kresse_1994,Kresse_1996a}, employing the projector augmented-wave (PAW) formalism~\cite{Blochl_1994,Kresse_1996b} to treat core electrons and a plane-wave basis set for the valence electrons. Exchange-correlation effects were described by the meta-GGA r$^2$SCAN functional~\cite{Furness_2020}, which has demonstrated reliable accuracy for the structural, electronic, and vibrational properties of deep-level defects in 4H-SiC~\cite{Abbas_2025,Zalandauskas_2025,Younesi_2026}. Defect supercells were constructed from the hexagonal 4H-SiC primitive cell using a $6\times6\times2$ supercell with 576 atomic sites. Brillouin zone integration was done at the $\Gamma$ point, and the kinetic energy cutoff for the plane-wave basis was set to $600$~eV. Total energies and ionic forces were converged to $10^{-8}$~eV and \SI{0.1}{\milli\electronvolt\per\angstrom}, respectively.

The transition dipole moment ($\boldsymbol{\mu}_{ij}$, TDM) between states $i$ and $j$ are given by
\begin{equation}
    \boldsymbol{\mu}_{ij} = \left\langle \psi_i \right| q\, \boldsymbol{r} \left| \psi_j \right\rangle ,
\end{equation}
where $q$ is the electron charge and $\boldsymbol{r}$ is the position operator. The Slater--Condon rule allows the transition dipole moment to be evaluated using Kohn--Sham single-particle states, $\psi_i$ and $\psi_j$, rather than the many-electron wavefunctions, $\Psi_i$ and $\Psi_j$. In this work, $\psi_i$ and $\psi_j$ are constructed from single-particle Kohn--Sham orbitals obtained from the triplet electronic configuration, which is a standard approximation in density functional theory calculations~\cite{Razinkovas_2021_photoionization,Davidsson2020_TDM,Maciaszek_2024_blue,Yan_2025,Meher_2025}.

\subsection{Temperature-dependent PL spectroscopy}
The PL temperature series were performed with a photoluminescence setup consisting of a microscope (10x objective, Mitutoyo) coupled to an imaging spectrograph (Horriba Jobin Yvon, iHR550) equipped with 300 grooves/mm gratings. The spectrometer was coupled to an InGaAs detector array (Andor DU491A), providing a spectral resolution of $\sim$0.14~nm with the 300~gr/mm grating. The sample was initially cooled to cryogenic temperatures ($\sim$3.5~K) using a closed-cycle He cryostat and measured at different temperatures using a temperature controller (Oxford Instruments). 
A 852~nm continuous wave (cw) laser (power density of $\sim0.18$~kW/cm$^2$ in the beam spot) 
was used to optically excite the sample. A long-pass filter (LP950) 
was used to suppress scattered excitation light. The laser excitation is directed toward the sample surface at an incident angle of 27$^\circ$ relative to the surface normal. 

\section*{Acknowledgments}
We acknowledge support from NSF CAREER (Award 2047564) and AFOSR Young Investigator Program (Award FA9550-23-1-0266). This work was supported in part by the AFOSR DURIP Award Agreement No. FA9550-25-1-0111. Funding was provided by  Work at the Molecular Foundry was supported by the Office of Science, Office of Basic Energy Sciences, of the U.S. Department of Energy under Contract No. DE-AC02-05CH11231. Part of this study was carried out at the UC Davis Center for NanoMicro Manufacturing (CNM2).
Financial support was kindly provided by Akademiaavtalen between Equinor and the University of Oslo through the research project QSenS, and by the Research council of Norway through the Centre for Defects in Semiconductors for Quantum Sensing (Project No.~354831) and the Norwegian Micro- and Nano-Fabrication Facility, NorFab, Project No.~349807. 
The computations were performed on resources provided by UNINETT Sigma2 --- the National Infrastructure for High Performance Computing and Data Storage in Norway.

\bibliography{ref}

@article{Kresse_1993,
  title={Ab initio molecular dynamics for liquid metals},
  author={Kresse, Georg and Hafner, J{\"u}rgen},
  journal={Physical Review B},
  volume={47},
  number={1},
  pages={558--561},
  year={1993},
  doi={10.1103/physrevb.47.558}
}

@ARTICLE{Ziegler2010,
  author = {J. F. Ziegler and M.D. Ziegler and J.P. Biersack},
  title = {{SRIM} {\textendash} The stopping and range of ions in matter (2010)},
  journal = {Nuclear Instruments and Methods in Physics Research Section B: Beam
	Interactions with Materials and Atoms},
  year = {2010},
  volume = {268},
  pages = {1818--1823},
  number = {11-12},
  month = {jun},
  doi = {10.1016/j.nimb.2010.02.091},
  publisher = {Elsevier {BV}}
}

@article{radulaski2017scalable,
  title={Scalable quantum photonics with single color centers in silicon carbide},
  author={Radulaski, Marina and Widmann, Matthias and Niethammer, Matthias and Zhang, Jingyuan Linda and Lee, Sang-Yun and Rendler, Torsten and Lagoudakis, Konstantinos G and Son, Nguyen Tien and Janzen, Erik and Ohshima, Takeshi and others},
  journal={Nano letters},
  volume={17},
  number={3},
  pages={1782--1786},
  year={2017},
  publisher={ACS Publications}
}

@article{Kresse_1994,
  title={Ab initio molecular-dynamics simulation of the liquid-metal{\textendash}amorphous-semiconductor transition in germanium},
  author={Kresse, Georg and Hafner, J{\"u}rgen},
  journal={Physical Review B},
  volume={49},
  number={20},
  pages={14251--14269},
  year={1994},
  doi={10.1103/physrevb.49.14251}
}

@article{Kresse_1996a,
  title={Efficiency of ab-initio total energy calculations for metals and semiconductors using a plane-wave basis set},
  author={Kresse, Georg and Furthm{\"u}ller, J{\"u}rgen},
  journal={Computational Materials Science},
  volume={6},
  number={1},
  pages={15--50},
  year={1996},
  doi={10.1016/0927-0256(96)00008-0}
}

@article{Blochl_1994,
  title={Projector augmented-wave method},
  author={Bl{\"o}chl, Peter E},
  journal={Physical Review B},
  volume={50},
  number={24},
  pages={17953},
  year={1994},
  doi={10.1103/physrevb.50.17953}
}

@article{Kresse_1996b,
  title={Efficient iterative schemes for ab initio total-energy calculations using a plane-wave basis set},
  author={Kresse, Georg and Furthm{\"u}ller, J{\"u}rgen},
  journal={Physical Review B},
  volume={54},
  number={16},
  pages={11169},
  year={1996},
  doi={10.1103/physrevb.54.11169}
}

@article{Furness_2020,
  title={Accurate and numerically efficient {r$^2$SCAN} meta-generalized gradient approximation},
  author={Furness, James W and Kaplan, Aaron D and Ning, Jinliang and Perdew, John P and Sun, Jianwei},
  journal={The Journal of Physical Chemistry Letters},
  volume={11},
  number={19},
  pages={8208--8215},
  year={2020},
  doi={10.1021/acs.jpclett.0c02405}
}

@article{Abbas_2025,
  title={Theoretical characterization of {NV}-like defects in {4H}-{SiC} using {ADAQ} with {SCAN} and {r$^2$SCAN} meta-{GGA} functionals},
  author={Abbas, Ghulam and Bulancea-Lindvall, Oscar and Davidsson, Joel and Armiento, Rickard and Abrikosov, Igor A},
  journal={Applied Physics Letters},
  volume={126},
  number={15},
  pages={154001},
  year={2025},
  doi={10.1063/5.0252129}
}

@article{Zalandauskas_2025,
  title={Theory of the divacancy in {4H}-{SiC}: impact of {J}ahn-{T}eller effect on optical properties},
  author={{\v{Z}}alandauskas, Vytautas and Silkinis, Rokas and Vines, Lasse and Razinkovas, Lukas and Bathen, Marianne Etzelm{\"u}ller},
  journal={npj Computational Materials},
  volume={11},
  number={1},
  pages={155},
  year={2025},
  doi={10.1038/s41524-025-01609-2}
}

@article{Norman2025,
  author={Norman, Victoria A.
  and Majety, Sridhar
  and Rubin, Alex H.
  and Saha, Pranta
  and Gonzalez, Nathan R.
  and Simo, Jeanette
  and Palomarez, Bradi
  and Li, Liang
  and Curro, Pietra B.
  and Dhuey, Scott
  and Virasawmy, Selven
  and Radulaski, Marina},
  title={Sub-2 Kelvin Characterization of Nitrogen-Vacancy Centers in Silicon Carbide   Nanopillars},
  journal={ACS Photonics},
  year={2025},
  month={May},
  day={21},
  publisher={American Chemical Society},
  volume={12},
  number={5},
  pages={2604-2611},
  doi={10.1021/acsphotonics.5c00096}
}

@article{Razinkovas_2021_photoionization,
  title={Photoionization of negatively charged {NV} centers in diamond: Theory and ab initio calculations},
  author={Razinkovas, Lukas and Maciaszek, Marek and Reinhard, Friedemann and Doherty, Marcus W and Alkauskas, Audrius},
  journal={Physical Review B},
  volume={104},
  number={23},
  pages={235301},
  year={2021},
  doi={10.1103/PhysRevB.104.235301}
}

@article{Davidsson2020_TDM,
  title={Theoretical polarization of zero phonon lines in point defects},
  author={Davidsson, Joel},
  journal={Journal of Physics: Condensed Matter},
  volume={32},
  number={38},
  pages={385502},
  year={2020},
  doi={10.1088/1361-648X/ab94f4}
}

@article{Maciaszek_2024_blue,
  title={Blue quantum emitter in hexagonal boron nitride and a carbon chain tetramer: a first-principles study},
  author={Maciaszek, Marek and Razinkovas, Lukas},
  journal={ACS Applied Nano Materials},
  volume={7},
  number={16},
  pages={18979--18985},
  year={2024},
  doi={10.1021/acsanm.4c02722}
}

@article{Yan_2025,
  title={Intralevel Optical Transitions of {XV} ({XV}= {BV}, {SiV}, and {NV}) Centers in Fluorinated Diamane},
  author={Yan, Longbin and Cheng, Shaobo and Ku, Yalun and Wang, Dongyang and Liu, Taiqiao and Li, Xing and Zhang, Zhaofu and Shan, Chongxin},
  journal={Nano Letters},
  volume={25},
  number={12},
  pages={4818--4824},
  year={2025},
  doi={10.1021/acs.nanolett.4c06343}
}

@article{Meher_2025,
  title={High-throughput computational search for group-{IV}-related quantum defects as spin-photon interfaces in {4H-SiC}},
  author={Meher, Shibu and Dey, Manoj and Singh, Abhishek Kumar},
  journal={Physical Review B},
  volume={112},
  number={18},
  pages={184112},
  year={2025},
  doi={10.1103/lsxj-nvhw}
}

@article{Shafizadeh_2024,
  title={Selection rules in the excitation of the divacancy and the nitrogen-vacancy pair in {4H-} and {6H-SiC}},
  author={Shafizadeh, Danial and Davidsson, Joel and Ohshima, Takeshi and Abrikosov, Igor A and Son, Nguyen T and Ivanov, Ivan G},
  journal={Physical Review B},
  volume={109},
  number={23},
  pages={235203},
  year={2024},
  doi={10.1103/PhysRevB.109.235203}
}

@article{Stenlund_2025,
  title={{ADAQ-SYM}: Automated symmetry analysis of defect orbitals},
  author={Stenlund, William and Davidsson, Joel and Armiento, Rickard and Iv{\'a}dy, Viktor and Abrikosov, Igor A},
  journal={Computer Physics Communications},
  volume={308},
  pages={109468},
  year={2025},
  doi={10.1016/j.cpc.2024.109468}
}

@article{Younesi_2026,
  title={Unraveling the electronic structure of silicon vacancy centers in {4H}-{SiC}},
  author={Ali Tayefeh Younesi and Minh Tuan Luu and Christopher Linderälv and Vytautas Žalandauskas and Marianne Etzelmüller Bathen and Nguyen Tien Son and Takeshi Ohshima and Gergő Thiering and Lukas Razinkovas and Ronald Ulbricht},
  journal={arXiv preprint arXiv:2602.14818},
  url={https://arxiv.org/abs/2602.14818},
  year={2026}
}

@article{PhysRevB.98.214113,
  title = {Electron paramagnetic resonance tagged high-resolution excitation spectroscopy of {NV}-centers in {4H-SiC}},
  author = {Zargaleh, S. A. and von Bardeleben, H. J. and Cantin, J. L. and Gerstmann, U. and Hameau, S. and Ebl\'e, B. and Gao, Weibo},
  journal = {Phys. Rev. B},
  volume = {98},
  issue = {21},
  pages = {214113},
  numpages = {8},
  year = {2018},
  month = {Dec},
  publisher = {American Physical Society},
  doi = {10.1103/PhysRevB.98.214113}
}

@article{PhysRevLett.124.223601,
  title = {Coherent Control of Nitrogen-Vacancy Center Spins in Silicon Carbide at Room Temperature},
  author = {Wang, Jun-Feng and Yan, Fei-Fei and Li, Qiang and Liu, Zheng-Hao and Liu, He and Guo, Guo-Ping and Guo, Li-Ping and Zhou, Xiong and Cui, Jin-Ming and Wang, Jian and Zhou, Zong-Quan and Xu, Xiao-Ye and Xu, Jin-Shi and Li, Chuan-Feng and Guo, Guang-Can},
  journal = {Phys. Rev. Lett.},
  volume = {124},
  issue = {22},
  pages = {223601},
  numpages = {6},
  year = {2020},
  month = {Jun},
  publisher = {American Physical Society},
  doi = {10.1103/PhysRevLett.124.223601}
}

@Article{Mu2020,
    author={Mu, Zhao
    and Zargaleh, Soroush Abbasi
    and von Bardeleben, Hans J{\"u}rgen
    and Fr{\"o}ch, Johannes E.
    and Nonahal, Milad
    and Cai, Hongbing
    and Yang, Xinge
    and Yang, Jianqun
    and Li, Xingji
    and Aharonovich, Igor
    and Gao, Weibo},
    title={Coherent Manipulation with Resonant Excitation and Single Emitter Creation of Nitrogen Vacancy Centers in {4H} Silicon Carbide},
    journal={Nano Letters},
    year={2020},
    month={Aug},
    day={12},
    publisher={American Chemical Society},
    volume={20},
    number={8},
    pages={6142-6147},
    issn={1530-6984},
    doi={10.1021/acs.nanolett.0c02342}
}

@article{PhysRevB.94.060102,
  title = {Evidence for near-infrared photoluminescence of nitrogen vacancy centers in {4H-SiC}},
  author = {Zargaleh, S. A. and Eble, B. and Hameau, S. and Cantin, J.-L. and Legrand, L. and Bernard, M. and Margaillan, F. and Lauret, J.-S. and Roch, J.-F. and von Bardeleben, H. J. and Rauls, E. and Gerstmann, U. and Treussart, F.},
  journal = {Phys. Rev. B},
  volume = {94},
  issue = {6},
  pages = {060102},
  numpages = {5},
  year = {2016},
  month = {Aug},
  publisher = {American Physical Society},
  doi = {10.1103/PhysRevB.94.060102}
}

@article{PhysRevB.92.064104,
  title = {Identification and magneto-optical properties of the {NV} center in {4H-SiC}},
  author = {von Bardeleben, H. J. and Cantin, J. L. and Rauls, E. and Gerstmann, U.},
  journal = {Phys. Rev. B},
  volume = {92},
  issue = {6},
  pages = {064104},
  numpages = {6},
  year = {2015},
  month = {Aug},
  publisher = {American Physical Society},
  doi = {10.1103/PhysRevB.92.064104}
}

@misc{zhigulin2026couplingnitrogenvacancycenters,
      title={Coupling nitrogen vacancy centers in silicon carbide to nanophotonic resonators}, 
      author={Ivan Zhigulin and Konosuke Shimazaki and Samuel M. Stephens and Angus Gale and Karin Yamamura and Hark Hoe Tan and Igor Aharonovich and Mehran Kianinia},
      year={2026},
      eprint={2602.21505},
      archivePrefix={arXiv},
      primaryClass={physics.optics},
      url={https://arxiv.org/abs/2602.21505}, 
}

@article{koehl_room_2011,
	title = {Room temperature coherent control of defect spin qubits in silicon carbide},
	volume = {479},
	copyright = {2011 Springer Nature Limited},
	issn = {1476-4687},
	doi = {10.1038/nature10562},
	language = {en},
	number = {7371},
	urldate = {2025-05-14},
	journal = {Nature},
	author = {Koehl, William F. and Buckley, Bob B. and Heremans, F. Joseph and Calusine, Greg and Awschalom, David D.},
	month = nov,
	year = {2011},
	Publisher = {Nature Publishing Group},
	pages = {84--87},
}

@article{christle_isolated_2015,
	title = {Isolated electron spins in silicon carbide with millisecond coherence times},
	volume = {14},
	copyright = {2014 Springer Nature Limited},
	issn = {1476-4660},
	doi = {10.1038/nmat4144},
	language = {en},
	number = {2},
	urldate = {2026-04-09},
	journal = {Nature Materials},
	publisher = {Nature Publishing Group},
	author = {Christle, David J. and Falk, Abram L. and Andrich, Paolo and Klimov, Paul V. and Hassan, Jawad Ul and Son, Nguyen T. and Janzén, Erik and Ohshima, Takeshi and Awschalom, David D.},
	month = feb,
	year = {2015},
	pages = {160--163},
}

@article{anderson_five-second_2022,
	title = {Five-second coherence of a single spin with single-shot readout in silicon carbide},
	volume = {8},
	doi = {10.1126/sciadv.abm5912},
	number = {5},
	urldate = {2026-04-09},
	journal = {Science Advances},
	publisher = {American Association for the Advancement of Science},
	author = {Anderson, Christopher P. and Glen, Elena O. and Zeledon, Cyrus and Bourassa, Alexandre and Jin, Yu and Zhu, Yizhi and Vorwerk, Christian and Crook, Alexander L. and Abe, Hiroshi and Ul-Hassan, Jawad and Ohshima, Takeshi and Son, Nguyen T. and Galli, Giulia and Awschalom, David D.},
	month = feb,
	year = {2022},
	pages = {eabm5912},
}

@article{bathen_resolving_2021,
	title = {Resolving {Jahn}-{Teller} induced vibronic fine structure of silicon vacancy quantum emission in silicon carbide},
	volume = {104},
	doi = {10.1103/PhysRevB.104.045120},
	number = {4},
	urldate = {2025-11-13},
	journal = {Physical Review B},
	publisher = {American Physical Society},
	author = {Bathen, Marianne Etzelmüller and Galeckas, Augustinas and Karsthof, Robert and Delteil, Aymeric and Sallet, Vincent and Kuznetsov, Andrej Yu. and Vines, Lasse},
	month = jul,
	year = {2021},
	pages = {045120},
}

@article{Shafizadeh_2025,
  title={Evolution of the optically detected magnetic resonance spectra of divacancies in {4H-SiC} from liquid-helium to room temperature},
  author={Shafizadeh, Danial and Son, Nguyen T and Abrikosov, Igor A and Ivanov, Ivan G},
  journal={Physical Review B},
  volume={111},
  number={16},
  pages={165201},
  year={2025},
  doi={10.1103/PhysRevB.111.165201}
}

@Article{Wang2020,
    author={Wang, Jun-Feng
    and Liu, Zheng-Hao
    and Yan, Fei-Fei
    and Li, Qiang
    and Yang, Xin-Ge
    and Guo, Liping
    and Zhou, Xiong
    and Huang, Wei
    and Xu, Jin-Shi
    and Li, Chuan-Feng
    and Guo, Guang-Can},
    title={Experimental Optical Properties of Single Nitrogen Vacancy Centers in Silicon Carbide at Room Temperature},
    journal={ACS Photonics},
    year={2020},
    month={Jul},
    day={15},
    publisher={American Chemical Society},
    volume={7},
    number={7},
    pages={1611-1616},
    doi={10.1021/acsphotonics.0c00218}
}

@article{falk_polytype_2013,
    title = {Polytype control of spin qubits in silicon carbide},
    volume = {4},
    copyright = {2013 The Author(s)},
    issn = {2041-1723},
    url = {https://www.nature.com/articles/ncomms2854},
    doi = {10.1038/ncomms2854},
    language = {en},
    number = {1},
    urldate = {2026-03-13},
    journal = {Nature Communications},
    publisher = {Nature Publishing Group},
    author = {Falk, Abram L. and Buckley, Bob B. and Calusine, Greg and Koehl, William F. and Dobrovitski, Viatcheslav V. and Politi, Alberto and Zorman, Christian A. and Feng, Philip X.-L. and Awschalom, David D.},
    month = may,
    year = {2013},
    pages = {1819},
}

@article{son_modified_2022,
	title = {Modified divacancies in {4H}-{SiC}},
	volume = {132},
	issn = {0021-8979},
	url = {https://doi.org/10.1063/5.0099017},
	doi = {10.1063/5.0099017},
	number = {2},
	urldate = {2025-10-06},
	journal = {Journal of Applied Physics},
	author = {Son, N. T. and Shafizadeh, D. and Ohshima, T. and Ivanov, I. G.},
	month = jul,
	year = {2022},
	pages = {025703},
}

@article{Shibata_1998,
    doi = {10.1143/JJAP.37.550},
    url = {https://doi.org/10.1143/JJAP.37.550},
    year = {1998},
    month = {feb},
    publisher = {},
    volume = {37},
    number = {2R},
    pages = {550},
    author = {Shibata, Hajime},
    title = {Negative Thermal Quenching Curves in Photoluminescence of Solids},
    journal = {Japanese Journal of Applied Physics},
}

@article{PhysRevB.82.115207,
    title = {Excited state properties of donor bound excitons in ZnO},
    author = {Meyer, Bruno K. and Sann, Joachim and Eisermann, Sebastian and Lautenschlaeger, Stefan and Wagner, Markus R. and Kaiser, Martin and Callsen, Gordon and Reparaz, Juan S. and Hoffmann, Axel},
    journal = {Phys. Rev. B},
    volume = {82},
    issue = {11},
    pages = {115207},
    numpages = {8},
    year = {2010},
    month = {Sep},
    publisher = {American Physical Society},
    doi = {10.1103/PhysRevB.82.115207},
    url = {https://link.aps.org/doi/10.1103/PhysRevB.82.115207}
}

@incollection{ivanov2023near,
    author    = {Ivanov, Ivan G. and Son, Nguyen T.},
    title     = {Near-Infrared Luminescent Centers in Silicon Carbide},
    editor    = {Feng, Zhe Chuan},
    booktitle = {Handbook of Silicon Carbide Materials and Devices},
    year      = {2023},
    edition   = {1st},
    chapter   = {10},
    pages     = {249--291},
    publisher = {CRC Press},
    address   = {Boca Raton, FL},
    doi       = {10.1201/9780429198540-13},
    url       = {https://www.taylorfrancis.com/chapters/mono/10.1201/9780429198540-13/near-infrared-luminescent-centers-silicon-carbide-ivan-ivanov-nguyen-son}
}

\clearpage

\begin{appendices}

\begin{center}
{\large \textbf{Supplementary Information for ``Seeing the forbidden: overcoming 
optical selection rules through nanophotonic integration''}}
\end{center}

\setcounter{section}{0}
\setcounter{figure}{0}
\setcounter{table}{0}
\setcounter{equation}{0}

\renewcommand{\thesection}{S\arabic{section}}
\renewcommand{\thesubsection}{S\arabic{section}.\arabic{subsection}}
\renewcommand{\theequation}{S\arabic{equation}}
\renewcommand{\thefigure}{S\arabic{figure}}
\renewcommand{\thetable}{S\arabic{table}}

\makeatletter
\def\@seccntformat#1{\csname the#1\endcsname.\quad}
\makeatother

\section{PL temperature series of VV$^0$ and NV$^{-}$ signals in 4H-SiC}
\begin{figure*}[h]
    \centering
    \includegraphics[width=\linewidth]{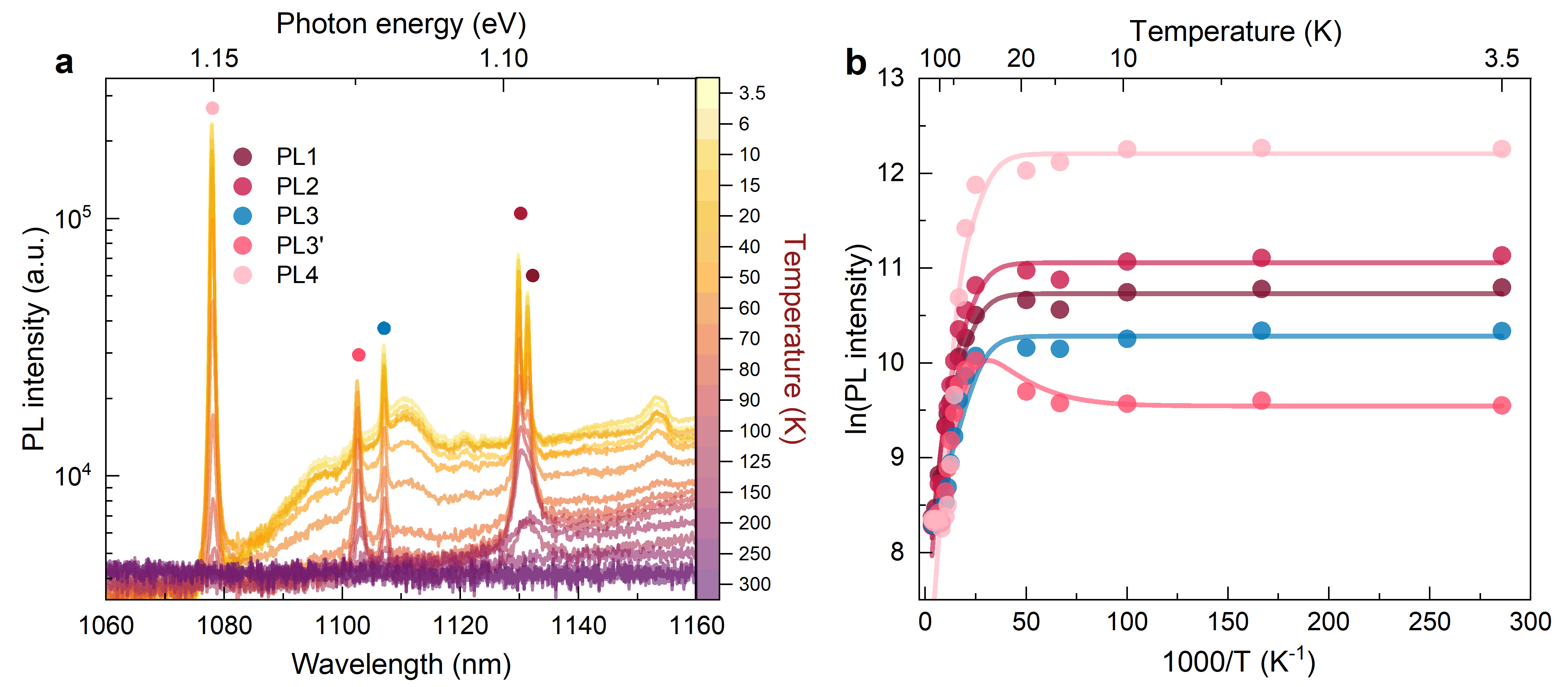}
    \caption{\textbf{PL temperature series of VV$^{0}$.} (a) PL spectra of divacancy signals PL1-PL4, including PL3$'$ measured at temperatures from 3.5~K (light yellow curve) to 300~K (dark purple curve). Excitation was conducted at 852~nm. (b) Arrhenius plots of all divacancy signals PL1-PL4, including PL3$'$.}
    \label{fig:pl1-4tempserie}
\end{figure*}

\begin{figure*}[h]
    \centering
    \includegraphics[width=\linewidth]{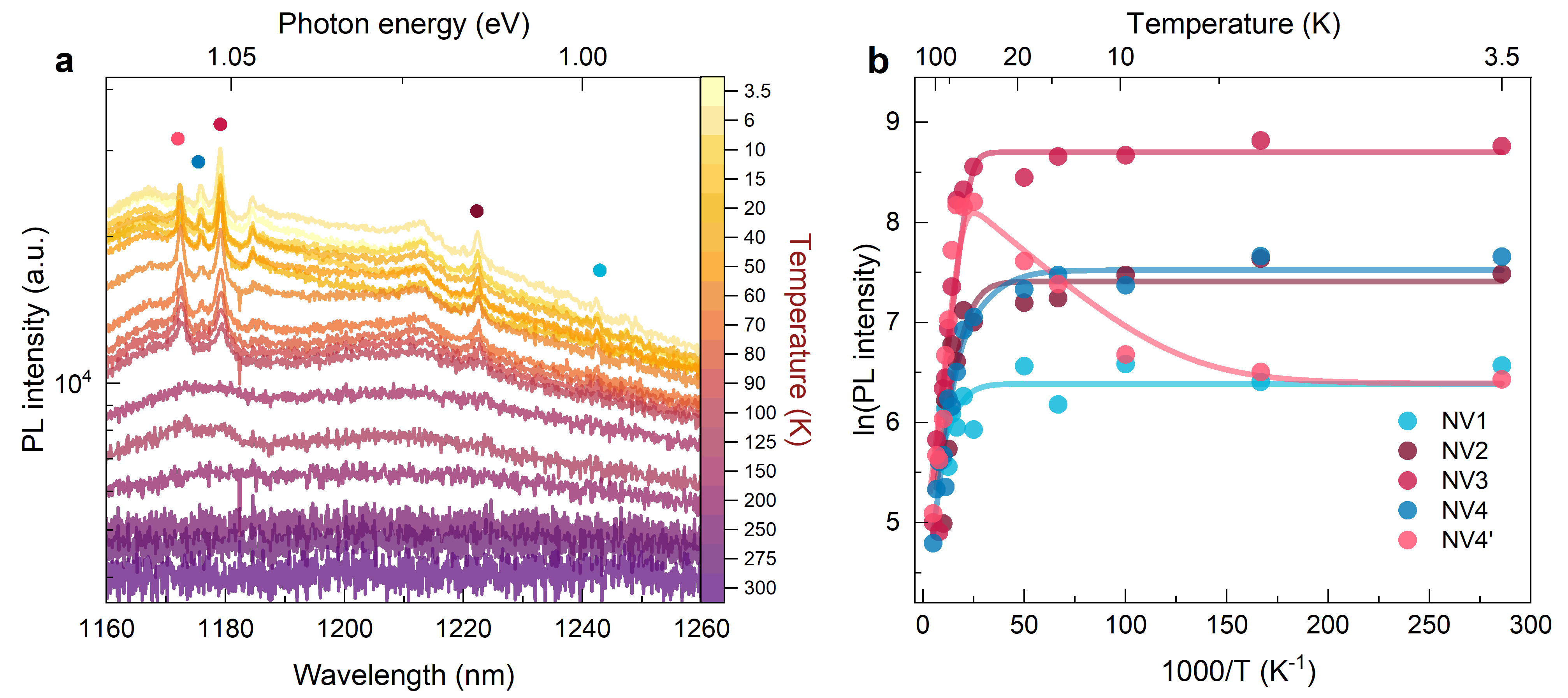}
    \caption{\textbf{PL temperature series of NV$^{-}$.} (a) PL spectra of NV signals NV1-4, including NV4$'$ measured at temperatures from 3.5~K (light yellow curve) to 300~K (dark purple curve). Excitation was conducted at 852~nm. (b) Arrhenius plots of all NV signals NV1-4, including NV4$'$. The amplitude of the PL intensity on the $y$-axis of the Arrhenius plot is the maximum of the fitted Lorentzian peak of the baseline-corrected PL signal.}
    \label{fig:nv1-4tempserie}
\end{figure*}

A PL temperature series of all divacancy signals (PL1, PL2, PL3, PL3$'$, and PL4) is shown in Fig.~\ref{fig:pl1-4tempserie}\textbf{a}. The peaks are marked with circles with distinct colors, corresponding to the Arrhenius plots for each peak in Fig.~\ref{fig:pl1-4tempserie}\textbf{b}. 
Similarly, for the NV-signals (NV1, NV2, NV3, NV4 and NV4'), a PL temperature series is shown in Fig.~\ref{fig:nv1-4tempserie}\textbf{a}, where the peaks are marked with circles of distinct colors that correspond to the Arrhenius plots in Fig.~\ref{fig:nv1-4tempserie}\textbf{b}. The experimental data is fitted with asymptotic curves, with fitting parameters found by linear regression. 
Note that the PL temperature series are collected from n-type 4H-SiC samples without nanopillars, see Section~\ref{sec:sample_prep}. 

Arrhenius curves reveal which thermally activated processes reign in different temperature regimes. This accounts for thermal activation of defects in solids, among other phenomena. In the Arrhenius plots, the logarithm of the PL intensity as a function of the inverse temperature can be fitted with an asymptotic curve, given by
\begin{equation}
    I(T)=I_0\frac{1+\Sigma_iA_i e^{\frac{-E_i}{k_BT}}}{1+\Sigma_jA_j e^{\frac{-E_j}{k_BT}}},
    \label{eq:asymptotic}
\end{equation}
where $E_i$ and $E_j$ are the activation energies for the different thermal processes, i.e., negative and total thermal quenching, respectively.   
The fitting parameters (activation energies) for the asymptotic fits, here denoted $E_0$, $E_1$ and $E_2$, are found by linearly fitting separate regions of the experimental data. 
The activation energies correspond to the slope of the linear fit. 
In this case, we consider two thermal quenching asymptotes for the peaks without negative thermal quenching ($E_1$ and $E_2$) and three asymptotes for the peaks exhibiting negative thermal quenching ($E_0$, $E_1$, and $E_2$). The fitting function for the latter case, based on Eq.~\ref{eq:asymptotic}, becomes 
\begin{equation}
    I(T)=I_0\frac{1+A_{i,0} e^{\frac{-E_0}{k_BT}}}{1+A_{j,0} e^{\frac{-E_0}{k_BT}}+A_{j,1} e^{\frac{-E_1}{k_BT}}+A_{j,2} e^{\frac{-E_2}{k_BT}}} .
    \label{eq:asymptotic_specific}
\end{equation}

Visibly, all peaks PL1-4 in Fig.~\ref{fig:pl1-4tempserie}\textbf{b} and NV1-4 in Fig.~\ref{fig:nv1-4tempserie}\textbf{b} become quenched at elevated temperatures, initially with $E_1\sim16$~meV and ultimately with $E_2\sim58$~meV for PL1--4 and E$_2\sim35$ meV for NV1--4. All activation energies are listed in Table~\ref{tab:activationEnergies}. 
The PL3$'$ and NV4$'$ peaks are the only ones exhibiting notable thermal activation with $E_0\sim-2.6$~meV and $E_0\sim-1.6$~meV, respectively. This negative thermal quenching behavior has previously been observed for higher excited states of optically active defects, e.g. the $\mathrm{V_{Si}}$ in 4H-SiC \cite{bathen_resolving_2021}. 

\begin{table}[h]
\centering
\caption{ Activation energies used for asymptotic fitting derived by fitting linear regions of the Arrhenius plots.}\label{tab:activationEnergies}
\begin{tabular}{cccc}\toprule
\textbf{Peak} & \textbf{$E_0$ (meV)} & \textbf{$E_1$ (meV)} & \textbf{$E_2$ (meV)} \\
\midrule 
PL1--4 & & 16 & 58 \\\
PL3$'$ & -2.6 & 16 & 58 \\
\midrule
NV1--4 &  & 16 & 35 \\\
NV3$'$ & -1.6& 16 & 35 \\
 \botrule
\end{tabular}

\end{table}

\section{Selection rules for the \texorpdfstring{$C_{3v}$}{C3v} point group}

We discuss the selection rules for the $C_{3v}$ point group. First, a character table for the $C_{3v}$ point group is presented in Table~\ref{tab:C3v_characters}. The first three columns represent a class of symmetry operations: the identity operation $E$, threefold rotations around the $z$-axis $C_3(z)$, and reflections in a vertical mirror plane $\sigma_v$. The final column lists how linear basis functions transform according to each irreducible representation. Each row corresponds to an irreducible representation of the $C_{3v}$ point group ($A_1$, $A_2$, and $E$).

The characters are defined as the traces of the representation matrices associated with a given symmetry operation, and therefore quantify how basis functions transform under these operations. For one-dimensional irreducible representations ($A_1$ and $A_2$), the characters reduce to $1$ for symmetric and $-1$ for antisymmetric behavior under different symmetry operations. For the two-dimensional $E$ representation, the characters reflect the combined transformation of multiple basis functions. For example, a character of 2 indicates that both basis functions remain unchanged under the identity operation, whereas a character of 0 signifies that the functions are mixed by the symmetry operation in such a way that their contributions cancel in the trace of the matrix. 

\begin{table}[h]
\centering
\caption{Character table for the $C_{3v}$ point group}
\begin{tabular}{ccccc}\toprule
 & $E$ & $2C_{3}(z)$ & $3\sigma_v$ & Linear \\
\midrule
$A_{1}$ & 1 &  1 &  1 & $z$ \\
$A_{2}$ & 1 &  1 & -1 & \\
$E$     & 2 & -1 &  0 & $(x,y)$ \\
\botrule
\end{tabular}
\label{tab:C3v_characters}
\end{table}

From standard group-theoretical selection rules, an optical transition is allowed only if the representation of the transition dipole moment operator, $\Gamma_\mu$, contains the totally symmetric irreducible representation $A_1$. The representation of the transition dipole moment is given by
\begin{equation}
    \Gamma_\mu = \Gamma_i \otimes \Gamma_r \otimes \Gamma_f,
\end{equation}
where $\otimes$ denotes the direct product. Here, $\Gamma_i$ and $\Gamma_f$ are the irreducible representations associated with the initial and final electronic wavefunctions $\Psi_i$ and $\Psi_j$, respectively, while $\Gamma_r$ corresponds to the irreducible representation of the light polarization direction, as defined by the linear functions in Table~\ref{tab:C3v_characters}. The direct product table for the $C_{3v}$ point group is given in Table~\ref{tab:C3v_product}.

\begin{table}[h]
\centering
\caption{Product table for the $C_{3v}$ point group}
\begin{tabular}{cccc}\toprule
      & $A_1$ & $A_2$ & $E$ \\
\midrule
$A_1$ & $A_1$ & $A_2$ & $E$ \\
$A_2$ & $A_2$ & $A_1$ & $E$ \\
$E$   & $E$   & $E$   & $A_1 \oplus A_2 \oplus E$ \\
\botrule
\end{tabular}
\label{tab:C3v_product}
\end{table}

\begin{figure*}
    \centering
    \includegraphics[width=1.00\textwidth]{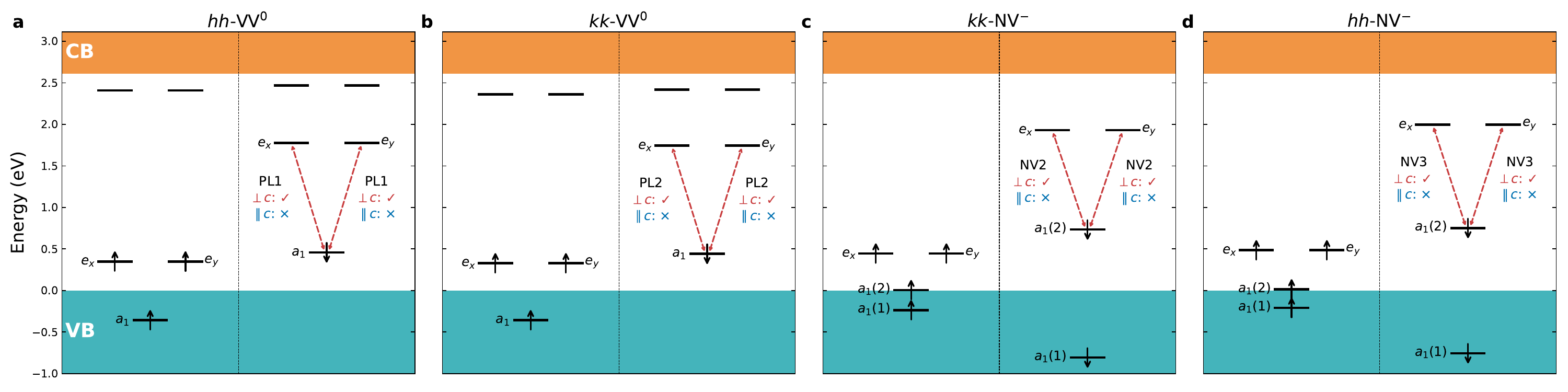}
    \caption{   
    \textbf{Electronic structure of axial defects.} Ground state Kohn--Sham single-particle defect level diagrams calculated at the r$^2$SCAN level of theory for the axial neutral divacancy ($\mathrm{V_{Si}V_C^{0}}$, VV$^0$) configurations $hh$ and $kk$ (\textbf{a}, \textbf{b}), and the axial negatively charged nitrogen-vacancy ($\mathrm{V_{Si}N_C^{-}}$, NV$^-$) configurations $kk$ and $hh$ (\textbf{c}, \textbf{d}) in 4H-SiC. Spin-majority (left) and spin-minority (right) channels are indicated by upward and downward arrows, respectively. Shaded regions denote the valence band (VB, teal) and conduction band (CB, orange). Dashed arrows indicate optically allowed transitions to excited states, with color encoding the polarization selection rules: red for transitions allowed under $\mathrm{E} \perp c$, blue for $\mathrm{E} \parallel c$, and both a red and a blue arrow shown side by side for transitions allowed under both polarizations.
    }
    \label{fig:sm_axial_KS}
\end{figure*}

The electronic structure of the axial ($hh$ and $kk$) neutral divacancy ($\mathrm{V_{Si}V_{C}^0}$ or VV$^0$) configurations can be described using six molecular orbitals derived from three carbon dangling bonds around the silicon vacancy site $\mathrm{V_{Si}}$ and three silicon dangling bonds around the carbon vacancy site $\mathrm{V_{C}}$. The three lowest-lying orbital levels (coming from carbon dangling bonds) within the band gap of 4H-SiC are an $a_1$ orbital and a higher lying set of doubly degenerate $e$ orbitals ($e_x$ and $e_y$). The ground state orbital occupation of the axial neutral divacancy is $a_1^2 e^2$. For the spin projection $m_s = 1$, this yields a single-determinant wave function $|a_1\,\bar{a}_1\,e_x\,e_y|$, where $|\cdots|$ denotes the Slater determinant and a bar over an orbital label (e.g., $\bar{a}_1$) denotes spin-down occupation, yielding a triplet ${^3\!A_2}$ state with $A_2$ symmetry. Promoting a spin-minority electron from the $a_1$ orbital to the doubly degenerate $e$ orbital produces the excited orbital configuration $a_1 e^3$ (see Fig.~\ref{fig:sm_axial_KS}\textbf{a},\textbf{b}). For $m_s = 1$, this configuration is spanned by two single-determinant wave functions, $|a_1\,e_x\,e_y\,\bar{e}_y|$ and $|a_1\,e_x\,\bar{e}_x\,e_y|$, which together form a degenerate excited state of triplet ${^3\!E}$ with $E$ symmetry.

The selection rules for  ${^3\!A_2} \leftrightarrow {^3\!E}$ transitions are as follows:
\begin{itemize}
    \item For $\mathrm{E} \perp c$, the dipole operator transforms as $(x, y)$, giving $\Gamma_r = E$. The direct product of $\Gamma_\mu$ is then:
    $$\Gamma_\mu = \Gamma_i \otimes \Gamma_r \otimes \Gamma_f = E \otimes E \otimes A_2 = A_1 \oplus A_2 \oplus E.$$ $\Gamma_\mu$ contains $A_1$, therefore the transition is allowed.

    \item For $\mathrm{E} \parallel c$, the dipole operator transforms as $z$, giving $\Gamma_r = A_1$. The direct product of $\Gamma_\mu$ is then:
    $$\Gamma_\mu = \Gamma_i \otimes \Gamma_r \otimes \Gamma_f = E \otimes A_1 \otimes A_2 = E.$$ $\Gamma_\mu$ does not contain $A_1$, therefore the transition is forbidden.
\end{itemize}

The electronic structure of the axial ($hh$ and $kk$) negatively charged nitrogen-vacancy ($\mathrm{V_{Si}N_{C}^-}$ or NV$^-$) configurations can be described by four molecular orbitals: two $a_1$ states and a doubly degenerate pair of $e$ states ($e_x$ and $e_y$). The lower-lying $a_1(1)$ orbital originates from the nitrogen dangling bond and lies deep within the valence band, while the upper $a_1(2)$ orbital and the degenerate $e$ states, derived from the carbon dangling bonds around the silicon vacancy site $\mathrm{V_{Si}}$, reside within the band gap of 4H-SiC (see Figs.~\ref{fig:sm_axial_KS}\textbf{c},\textbf{d}). Since $a_1(1)$ is fully occupied and energetically well separated from the gap states, the optically relevant ground state orbital occupation is $a_1(2)^2 e^2$. For the spin projection $m_s = 1$, this yields a single-determinant wave function $|a_1(2)\,\bar{a}_1(2)\,e_x\,e_y|$, giving a triplet ${^3\!A_2}$ state with $A_2$ symmetry. Promoting a spin-minority electron from the $a_1(2)$ orbital to the doubly degenerate $e$ orbital produces the excited orbital configuration $a_1(2)e^3$. For $m_s = 1$, this configuration is spanned by two single-determinant wave functions, $|a_1(2)\,e_x\,e_y\,\bar{e}_y|$ and $|a_1(2)\,e_x\,\bar{e}_x\,e_y|$, which together form a degenerate excited state of triplet $^3\!E$ with $E$ symmetry. The selection rules for the resulting ${^3\!A_2} \leftrightarrow {^3\!E}$ transition are therefore identical to those of the neutral axial divacancies, with $\mathrm{E} \perp c$ allowed and $\mathrm{E} \parallel c$ forbidden.

\section{Selection rules for the \texorpdfstring{$C_{1h}$}{C1h} point group \label{sec:c1h_selection}}

The character table for the $C_{1h}$ (also denoted as $C_s$) point group is given in Table~\ref{tab:C1h_characters}. The columns represent the identity operation $E$, reflection in a horizontal mirror plane $\sigma_h$, and the linear basis functions associated with each irreducible representation. For basal-plane defects in 4H-SiC, we take the horizontal mirror plane $\sigma_h$ to coincide with the $xz$-plane, where the $z$-axis is parallel to the crystallographic $c$-axis (the $[0001]$ direction). The two irreducible representations of the $C_{1h}$ point group are $A'$, consisting of functions symmetric with respect to $\sigma_h$, and $A''$, consisting of functions antisymmetric with respect to $\sigma_h$.

\begin{table}[h]
\centering
\caption{Character table for the $C_{1h}$ point group.}
\begin{tabular}{cccc}\toprule
      & $E$ & $\sigma_h$ & Linear \\
\midrule
$A'$  & 1 &  1 & $x,z$ \\
$A''$ & 1 & -1 & $y$ \\
\botrule
\end{tabular}
\label{tab:C1h_characters}
\end{table}

The same group-theoretical criterion applies to the $C_{1h}$ point group, where an optical transition is allowed only if the representation of the transition dipole moment operator, $\Gamma_\mu$, contains the totally symmetric irreducible representation (which is $A'$ for the $C_{1h}$ point group). The Cartesian components $x$ and $z$ transform as $A'$, while $y$ transforms as $A''$. Consequently, the irreducible representation of the light polarization, $\Gamma_{\mathrm{r}}$, transforms as $A'$ for polarization along the $c$-axis ($\mathrm{E} \parallel c$) and for in-plane polarization within the basal plane ($\mathrm{E} \perp c$), whereas out-of-plane polarization transforms as $A''$. The corresponding direct product table for the $C_{1h}$ point group is given in Table~\ref{tab:C1h_product}.

\begin{table}[h]
\centering
\caption{Product table for the $C_{1h}$ point group.}
\begin{tabular}{ccc}\toprule
      & $A'$  & $A''$ \\
\midrule
$A'$  & $A'$  & $A''$ \\
$A''$ & $A''$ & $A'$  \\
\botrule
\end{tabular}
\label{tab:C1h_product}
\end{table}

\begin{table}[h]
\centering
\caption{Optical selection rules for spin-triplet transitions in the $C_{1h}$ point group. The representation of the transition dipole moment $\Gamma_\mu$ is given by $\Gamma_\mu = \Gamma_i \otimes \Gamma_r \otimes \Gamma_f$, where $\Gamma_i$ and $\Gamma_f$ denote the irreducible representations of the initial and final electronic wavefunctions, $\Psi_i$ and $\Psi_f$, respectively. The term $\Gamma_r$ corresponds to the irreducible representation of the light polarization direction, as defined by the linear basis functions listed in Table~\ref{tab:C1h_characters}.}
\begin{tabular}{ccccc}\toprule
Transition & Polarization & $\Gamma_r$ & $\Gamma_\mu$ & Allowed \\
\midrule
\multirow{3}{*}{${^3\!A'} \leftrightarrow {^3\!A'}$} & $\mathrm{E} \perp c$ (in-plane) & $A'$  & $A' \otimes A' \otimes A' = A'$   & Yes \\
& $\mathrm{E} \perp c$ (out-of-plane) & $A''$ & $A' \otimes A'' \otimes A' = A''$ & No \\
& $\mathrm{E} \parallel c$            & $A'$  & $A' \otimes A' \otimes A' = A'$   & Yes \\
\midrule
\multirow{3}{*}{${^3\!A''} \leftrightarrow {^3\!A''}$} & $\mathrm{E} \perp c$ (in-plane) & $A'$  & $A'' \otimes A' \otimes A'' = A'$   & Yes \\
& $\mathrm{E} \perp c$ (out-of-plane) & $A''$ & $A'' \otimes A'' \otimes A'' = A''$ & No \\
& $\mathrm{E} \parallel c$            & $A'$  & $A'' \otimes A' \otimes A'' = A'$   & Yes \\
\midrule
\multirow{3}{*}{${^3\!A'} \leftrightarrow {^3\!A''}$}  & $\mathrm{E} \perp c$ (in-plane) & $A'$  & $A' \otimes A' \otimes A'' = A''$ & No \\
& $\mathrm{E} \perp c$ (out-of-plane) & $A''$ & $A' \otimes A'' \otimes A'' = A'$ & Yes \\
& $\mathrm{E} \parallel c$            & $A'$  & $A' \otimes A' \otimes A'' = A''$ & No \\
\botrule
\end{tabular}
\label{tab:C1h_selection_rules}
\end{table}

\begin{figure*}
    \centering
    \includegraphics[width=1.00\textwidth]{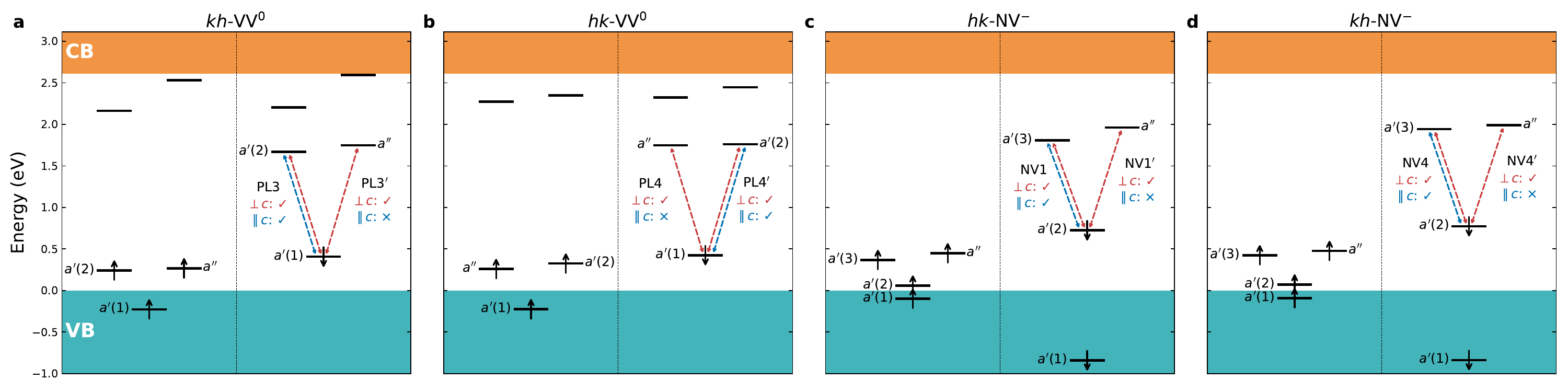}
    \caption{   
    \textbf{Electronic structure of basal defects.} Ground state Kohn--Sham single-particle defect level diagrams calculated at the r$^2$SCAN level of theory for the basal neutral divacancy ($\mathrm{V_{Si}V_C^{0}}$, VV$^0$) configurations $kh$ and $hk$ (\textbf{a}, \textbf{b}), and the basal negatively charged nitrogen-vacancy ($\mathrm{V_{Si}N_C^{-}}$, NV$^-$) configurations $hk$ and $kh$ (\textbf{c}, \textbf{d}) in 4H-SiC. Spin-majority (left) and spin-minority (right) channels are indicated by upward and downward arrows, respectively. Shaded regions denote the valence band (VB, teal) and conduction band (CB, orange). Dashed arrows indicate optically allowed transitions to excited states, with color encoding the polarization selection rules: red for transitions allowed under $\mathrm{E} \perp c$, blue for $\mathrm{E} \parallel c$, and both a red and a blue arrow shown side by side for transitions allowed under both polarizations.
    }
    \label{fig:sm_basal_KS}
\end{figure*}

\subsection{Basal configurations of the neutral divacancy }

The basal ($kh$ and $hk$) configurations of the neutral divacancy ($\mathrm{V_{Si}V_{C}^{0}}$, VV$^0$) exhibit reduced symmetry, which lifts the degeneracy of the $e$ states and splits them into non-degenerate levels of either $a'$ or $a''$ symmetry. The three lowest-lying defect orbitals within the 4H-SiC band gap, originating from carbon dangling bonds, are $a'(1)$, $a'(2)$, and $a''$ for the $kh$ configuration (ordered from lowest to highest energy, see Fig.~\ref{fig:sm_basal_KS}\textbf{a}), whereas for the $hk$ configuration the ordering becomes $a'(1)$, $a''$, and $a'(2)$ (see Fig.~\ref{fig:sm_basal_KS}\textbf{b}).

For the $kh$-VV$^0$ configuration, the ground state wave function for the $m_s=1$ spin projection is described by a single-determinant $|a'(1)\,\bar{a'}(1)\,a'(2)\,a''|$, corresponding to a triplet ${^3\!A''}$ state with $A''$ symmetry. The first excited state is obtained by promoting a spin-minority electron from the $a'(1)$ to the $a'(2)$ orbital, yielding the single-determinant wavefunction $|a'(1)\,a'(2)\,\bar{a'}(2)\,a''|$, which corresponds to a ${^3\!A''}$ state with $A''$ symmetry. The second excited state arises by promoting an electron from the $a'(1)$ to the $a''$ orbital in the spin-minority channel, resulting in the $|a'(1)\,a'(2)\,a''\,\bar{a''}|$ wavefunction and giving a ${^3\!A'}$ state with $A'$ symmetry. The corresponding optical selection rules for the PL3 ($kh$) line, associated with the ${^3\!A''} \rightarrow {^3\!A''}$ transition, and the PL3$'$ ($kh'$) line, associated with the ${^3\!A'} \rightarrow {^3\!A''}$ transition, are summarized in Table~\ref{tab:C1h_selection_rules} and illustrated by the red and blue dashed arrows in Fig.~\ref{fig:sm_basal_KS}\textbf{a}.

For the $hk$-VV$^0$ configuration, the ground state wavefunction for the $m_s=1$ spin projection can be represented by a single-determinant $|a'(1)\,\bar{a'}(1)\,a''\,a'(2)|$, corresponding to a triplet ${^3\!A''}$ state with $A''$ symmetry. The first excited state involves promotion of a spin-minority electron from the $a'(1)$ to the $a''$ orbital, resulting in the single-determinant wavefunction $|a'(1)\,a''\,\bar{a''}\,a'(2)|$ and a ${^3\!A'}$ state, while the second excited state corresponds to promotion from $a'(1)$ to $a'(2)$, yielding $|a'(1)\,a''\,a'(2)\,\bar{a'}(2)|$ and a ${^3\!A''}$ state. For the $hk$-VV$^0$ configuration, the optical selection rules corresponding to the PL4 ($hk$) line, associated with the ${^3\!A'} \rightarrow {^3\!A''}$ transition, and the PL4$'$ ($hk'$) line for the ${^3\!A''} \rightarrow {^3\!A''}$ transition, are given in Table~\ref{tab:C1h_selection_rules} and indicated by dashed arrows in Fig.~\ref{fig:sm_basal_KS}\textbf{b}. The experimental observation of the PL4$'$ line is reported in Figs.~6~and~7 of Ref.~\cite{Shafizadeh_2025}.

\subsection{Basal configurations of the negatively charged nitrogen-vacancy center}

The basal ($kh$ and $hk$) configurations of the negatively charged nitrogen-vacancy center ($\mathrm{V_{Si}N_{C}^{-}}$, NV$^-$) similarly exhibit reduced symmetry, lifting the degeneracy of the $e$ states and splitting them into non-degenerate $a'$ and $a''$ levels. The lowest-lying $a'(1)$ orbital, originating from a nitrogen dangling bond, is located deep within the valence band. The $a'(1)$ orbital is fully occupied and energetically well separated from the defect states in the band gap and can be neglected in the following discussion. The relevant in-gap states, derived from carbon dangling bonds, are $a'(2)$, $a'(3)$, and $a''$, ordered from lowest to highest energy (see Fig.~\ref{fig:sm_basal_KS}\textbf{c},\textbf{d}).

For both basal configurations, the ground state wavefunction for the $m_s=1$ spin projection can be represented by a single-determinant state $|a'(2)\,\bar{a'}(2)\,a'(3)\,a''|$, corresponding to a triplet  ${^3\!A''}$ state with $A''$ symmetry. The first excited state is obtained by promoting a spin-minority electron from $a'(2)$ to $a'(3)$ orbital, yielding $|a'(2)\,a'(3)\,\bar{a'}(3)\,a''|$ single-determinant wavefunction, which remains a ${^3\!A''}$ state. The second excited state arises by promoting a spin-minority electron from the $a'(2)$ orbital to the $a''$ orbital, resulting in $|a'(2)\,a'(3)\,a''\,\bar{a''}|$ single-determinant wave function and a ${^3\!A'}$ state with $A'$ symmetry. The optical selection rules for the NV1 ($hk$), NV1$'$ ($hk'$), NV4 ($kh$), and NV4$'$ ($kh'$) lines --- corresponding to the ${^3\!A''}\rightarrow{^3\!A''}$, ${^3\!A'}\rightarrow{^3\!A''}$, ${^3\!A''}\rightarrow{^3\!A''}$, and ${^3\!A'}\rightarrow{^3\!A''}$ transitions, respectively --- are summarized in Table~\ref{tab:C1h_selection_rules}. These selection rules are further illustrated by the red and blue dashed arrows in Fig.~\ref{fig:sm_basal_KS}\textbf{c},\textbf{d}.

\end{appendices}

\end{document}